\documentclass[aps,prl,superscriptaddress,reprint]{revtex4-2}
\usepackage{centernot}
\usepackage{graphicx}
\usepackage{amsmath}
\usepackage{times}
\usepackage{amssymb}
\usepackage{mathrsfs}
\usepackage{chemarr}
\usepackage{xcolor}
\usepackage{url}
\usepackage{version}
\usepackage{tikz}

\definecolor{linkcolor}{rgb}{0,0,0.6}
\usepackage[pdftex,colorlinks=true,pdfstartview=FitV,linkcolor= linkcolor,citecolor= linkcolor,urlcolor= linkcolor,hyperindex=true,hyperfigures=false]{hyperref}
\usepackage{siunitx}

\newcommand{\cD}{\mathcal{D}}

\newcommand\bfr{{\bf r}}
\newcommand\bfq{{\bf q}}
\newcommand\bfu{{\bf u}}

\newcommand\alb[1]{{\color{blue} #1}}

\newcommand\mfu{\mathfrak{u}}

\definecolor{darkgreen}{cmyk}{1,0,1,0.1}

\begin{document} 
\title{Non-reciprocity across scales in active mixtures}

\author{Alberto Dinelli}
\affiliation{Universit\'e Paris Cit\'e, Laboratoire Mati\`ere et Syst\`emes Complexes (MSC), UMR 7057 CNRS,  F-75205, 75205 Paris, France}
\author{J\'er\'emy O'Byrne}
\affiliation{Universit\'e Paris Cit\'e, Laboratoire Mati\`ere et Syst\`emes Complexes (MSC), UMR 7057 CNRS,  F-75205, 75205 Paris, France}
\affiliation{Department of Applied Maths and Theoretical Physics, University of Cambridge, Centre for Mathematical Sciences, Wilberforce Rd, Cambridge CB3 0WA, UK}
\author{Agnese Curatolo}
\affiliation{John A. Paulson School of Engineering and Applied Sciences and
  Kavli Institute for Bionano Science and Technology, Harvard
  University, Cambridge, MA 02138, USA}
\author{Yongfeng Zhao}
\affiliation{Center for Soft Condensed Matter Physics and Interdisciplinary Research \& School of Physical Science and Technology, Soochow University, 215006 Suzhou, China}
\author{Peter Sollich}
\affiliation{Institute for Theoretical Physics, Georg-August-Universität Göttingen, 37 077 Göttingen, Germany}
\affiliation{Department of Mathematics, King's College London, London WC2R 2LS, UK}
\author{Julien Tailleur}
\affiliation{Universit\'e Paris Cit\'e, Laboratoire Mati\`ere et Syst\`emes Complexes (MSC), UMR 7057 CNRS,  F-75205, 75205 Paris, France}
\affiliation{Department of Physics, Massachusetts Institute of Technology, Cambridge, Massachusetts 02139, USA}
\date{\today}



\begin{abstract}

  In active matter, the lack of momentum conservation makes
  non-reciprocal interactions the rule rather than the exception. They
  lead to a rich set of emerging behaviors that are hard to account
  for and to predict starting from the microscopic scale, due to the
  absence of a generic theoretical framework out of equilibrium. Here
  we consider bacterial mixtures that interact via mediated,
  non-reciprocal interactions like quorum-sensing and chemotaxis. By
  explicity relating microscopic and macroscopic dynamics, we show
  that non-reciprocity may fade as coarse-graining proceeds, leading to
  large-scale bona fide equilibrium descriptions. In turns, this
  allows us to account quantitatively, and without fitting parameters,
  for the rich behaviors observed in microscopic simulations including
  phase separation, demixing or multi-phase coexistence. We also
  derive the condition under which non-reciprocity is strong enough to
  survive coarse-graining, leading to a wealth of dynamical
  patterns. Again, the explicit coarse-graining of the dynamics allows
  us to predict the phase diagram of the system starting from its
  microscopic description. All in all, our work demonstrates that the
  fate of non-reciprocity across scales is a subtle and important
  question.
\end{abstract}

\maketitle

\section{Introduction}\label{sec1}

Our ability to design and engineer new materials largely relies on the
possibility to infer their large-scale properties from their
microscopic constituents. For equilibrium systems, statistical
mechanics allows us to do so by relating the macroscopic free energy
to the microscopic partition function and the Boltzmann weight. As a
result, the emerging properties of equilibrium systems can be
predicted by balancing energy and entropy. This general principle, at
the root of so many industrial innovations over the past century,
comes with a strong restriction: it only applies to the steady state
of systems satisfying detailed balance, thus excluding
the vast class of nonequilibrium systems and transient dynamical phenomena. An important challenge is thus to develop
theoretical frameworks that would allow us to relate the microscopic
description of nonequilibrium systems to their emerging behavior.

This is particularly important for active systems, which comprise
large assemblies of individual units able to exert non-conservative
forces on their environment~\cite{o2022time}. From spontaneously
flowing
matter~\cite{deseigne2010collective,schaller2010polar,sumino2012large,bricard_emergence_2013}
to living
crystals~\cite{theurkauff2012dynamic,palacci2013living,buttinoni2013dynamical,van_der_linden_interrupted_2019,tan2022odd},
active materials display phases without counterparts in equilibrium
physics~\cite{marchetti_hydrodynamics_2013}. This rich phenomenology
relies in part on the existence of non-reciprocal interactions (NRI)
between active particles, which have attracted a lot of attention
recently. From the spontaneous emergence of traveling waves to
anomalous mechanics and odd elasticity, NRI have indeed been shown to
lead to a wealth of exciting
phenomena~\cite{soto_self-assembly_2014,baek2018generic,saha_pairing_2019,agudo-canalejo_active_2019,saha_scalar_2020,you_nonreciprocity_2020,nasouri_exact_2020,granek2020bodies,ouazan-reboul_non-equilibrium_2021,fruchart_non-rec_2021,frohoff2021suppression,frohoff2021localized,poncet2022soft,gupta2022nonreciprocal}.

In the simplest case of systems with pairwise forces, NRI correspond
to the breakdown of Newton's third law, which states that if particle
$i$ exerts a force ${\bf f}_{ij}$ onto particle $j$ then ${\bf
  f}_{ji}=-{\bf f}_{ij}$. Active systems exchange momentum with their
environment and are thus free of this constraint\if{, which is
  enforced by momentum conservation}\fi. Note that pairwise forces are
an idealized limit for most active particles: experimental systems
instead typically involve complex mediated $N$-body interactions like
chemotaxis, quorum sensing, or hydrodynamic interactions. In all
cases, predicting how such microscopic interactions impact the
emerging behavior is a {challenging, indeed mostly impossible} task.
An appealing alternative has recently been proposed: to postulate
phenomenological theories in which action/reaction is directly broken
at the macroscopic
scale~\cite{saha_scalar_2020,you_nonreciprocity_2020}. The analysis of
the large-scale behavior then amounts to a non-linear dynamics problem
for which a wealth of tools are
available~\cite{cross1993pattern,saarloos_amplitude_1994,aranson2002world,rapp2019systematic,bergmann2018active,saha_scalar_2020,you_nonreciprocity_2020,frohoff2021suppression,frohoff2021localized}.
However, a major limitation is that, in the presence of NRI, there is
no generic way to infer which microscopic systems correspond to a
given macroscopic description. This not only prevents us from
assessing the scope of these theories, but it also deprives us of
guiding principles when it comes to engineering microscopic active
systems to realize the exciting emerging behaviors observed at the
macroscopic scale.

In this article, we bridge the gap between microscopic and macroscopic
descriptions of active systems with non-reciprocal interactions, which
allows us to show that the violation of action-reaction is strongly
scale dependent. To do so, we study active mixtures, which comprise
several types of interacting active particles and have attracted a lot
of interest
recently~\cite{soto_self-assembly_2014,stenhammar2015activity,yeo2015collective,wysocki2016propagating,wittkowski2017nonequilibrium,sturmer2019chemotaxis,saha_scalar_2020,you_nonreciprocity_2020,rodriguez2020phase,kolb2020active,bardfalvy2020symmetric,de2021active,fruchart_non-rec_2021,frohoff2021localized,frohoff2021suppression,de2021diversity,poncet2022soft,paoluzzi2020information,li2021hierarchical,williams2021confinement}.
We consider active particles that interact via quorum sensing (QS),
i.e.\ regulate their motility according to the local density of their
peers. QS is generic in nature~\cite{miller2001QS}, where
it is typically mediated by diffusing signalling molecules. For
microorganisms, it plays an important role in regulating diverse
biological functions, from
bioluminescence~\cite{nealson1970luminescence,engebrecht1984lux,fuqua1994quorum,verma2013quorum}
and virulence~\cite{tsou2010virulence} to biofilm
formation~\cite{hammer2003biofilm} and
swarming~\cite{daniels2004swarming}.  Furthermore, QS can also be
engineered in the lab, for instance using light-controlled
self-propelled
colloids~\cite{bauerle_self-organization_2018,lavergne_group_2019,massana-cid_rectification_2022}.

We consider $N$ `species' of active particles and denote by
$\rho_\mu(\bfr)$ the density field of species $\mu$. For concreteness,
we present our results for active Brownian particles (ABPs) and
run-and-tumble particles (RTPs), but we stress that they hold more
generally and also apply, for instance, to active Ornstein-Uhlenbeck
particles. QS interactions between the species then lead to the
following dynamics for particle $i$ of species $\mu$:
\begin{equation}\label{eq:microdyn}
   \dot \bfr_{i,\mu}=v_\mu(\bfr_i,[\{\rho_\nu\}]) \bfu_{i,\mu}\;,
\end{equation}
where the self-propulsion speed $v_\mu$ is both a function of $\bfr_i$
and a functional of all density fields.  The particle orientation
$\bfu_{i,\mu}$ is a unit vector {that} undergoes either rotational
diffusion (ABPs) or tumbles instantaneously (RTPs), with a
persistence time $\tau_\mu$~\footnote{Note that making $\tau_{i,\mu}$
  a function of $\bfr_i$ and a functional of $\{\rho_\nu\}$ does not
  lead to any interesting phenomenology so, for simplicity, we do not
  consider this case in the main text.}. We note that
Eq.~\eqref{eq:microdyn} is non-reciprocal by definition: the
displacements of particles $i$ and $j$ of species $\mu$ and $\nu$
impact their respective velocities in completely arbitrary ways. To
address how this non-reciprocity affects the large-scale properties,
we first coarse-grain the dynamics~\eqref{eq:microdyn}. We then
compute the entropy production rate of the resulting fluctuating
hyrodynamics and show that it \textit{vanishes} whenever
\begin{equation}\label{eq:intointegrable}
  \frac{\delta \log v_\mu(\bfr)}{\delta \rho_\nu(\bfr')} =
  \frac{\delta \log v_\nu(\bfr')}{\delta \rho_\mu(\bfr)}\;\qquad\text{for any\,}\mu, \nu.
\end{equation}
In this case, the microscopic non-equilibrium dynamics leads to an
effective large-scale equilibrium theory and non-reciprocity vanishes
upon coarse-graining. To the best of our knowledge,
condition~\eqref{eq:intointegrable} is the first non-trivial
generalization of Newton's action-reaction principle to a microscopic
model of active mixtures in the presence of many-body
mediated interactions. We show that the system then admits an
effective free energy which allows predicting its emerging
behavior. Remarkably, this allows us to construct the phase diagram of
the system from its microscopic dynamics without any fit parameters, a
rare achievement even in equilibrium. On the contrary, when
Eq.~\eqref{eq:intointegrable} is violated, non-reciprocity survives
coarse-graining and we derive a sufficient condition on the
\textit{microscopic} dynamics to observe travelling patterns that
explicitly break time-reversal symmetry at the macroscopic scale. All
in all, our work thus demonstrates that the fate of non-reciprocity
across scales
is a subtle and important question. To support the generality of this
statement, we close our article by extending our results to
chemotactic interactions where macroscopic reciprocity may again
emerge despite microscopic NRI.

\if{
In this article, we bridge the gap between microscopic and macroscopic
descriptions of active systems with non-reciprocal interactions, which
allows us to show that the violation of action-reaction is
strongly scale dependent. To do so, we study active mixtures, which
comprise several types of interacting active particles and have
attracted a lot of interest
recently~\cite{soto_self-assembly_2014,stenhammar2015activity,yeo2015collective,wysocki2016propagating,wittkowski2017nonequilibrium,sturmer2019chemotaxis,saha_scalar_2020,you_nonreciprocity_2020,rodriguez2020phase,kolb2020active,bardfalvy2020symmetric,de2021active,fruchart_non-rec_2021,frohoff2021localized,frohoff2021suppression,de2021diversity,poncet2022soft,paoluzzi2020information,li2021hierarchical,williams2021confinement}. {We consider active particles that interact via $N$-body, mediated interactions like quorum sensing (QS), which explicitly violate momentum conservation at the microscopic scale.} \\
{Quorum-sensing, that consists on the regulation of particle's motility based on the density of their peers}, is in fact more generic in nature~\cite{miller2001QS}, where it is typically mediated by diffusing signalling molecules. For microorganisms, it
plays an important role in regulating diverse biological functions,
from
bioluminescence~\cite{nealson1970luminescence,engebrecht1984lux,fuqua1994quorum,verma2013quorum}
and virulence~\cite{tsou2010virulence} to biofilm
formation~\cite{hammer2003biofilm} and
swarming~\cite{daniels2004swarming}.  Interestingly, QS can also be
engineered in the lab, for instance using light-controlled
self-propelled
colloids~\cite{bauerle_self-organization_2018,lavergne_group_2019,massana-cid_rectification_2022}.
We consider $N$ `species' of active particles and denote by
$\rho_\mu(\bfr)$ the density field of species $\mu$. \st{QS interactions
  between the species lead to the following dynamics for} \alb{The dynamics
  of particle $i$ of species $\mu$ will read}:
\begin{equation}\label{eq:microdyn}
   \dot \bfr_{i,\mu}=v_\mu(\bfr_i,[\{\rho_\nu\}]) \bfu_{i,\mu}\;,
\end{equation}
where $\bfu_{i,\mu}$ is the orientation of the particle and $v_\mu$ is the
self-propulsion speed. \st{generally both a function of $\bfr_i$
and a functional of all density fields.} For concreteness, we present
our results for active Brownian particles (ABPs) and run-and-tumble
particles (RTPs), even though our results also apply to active
Ornstein-Uhlenbeck particles.  Then, $\bfu_{i,\mu}$ is a unit vector
which undergoes either rotational diffusion (ABPs) or instantaneous
tumbles (RTPs), leading to a persistence time
$\tau_\mu$\if{~\footnote{Note that making $\tau_{i,\mu}$ a function of
$\bfr_i$ and a functional of $\{\rho_\nu\}$ does not lead to any
interesting phenomenology so that, for simplicity, we do not consider
this case in the main text.}}\fi. Finally, QS interactions between the particles are enforced as:
  \begin{equation}\label{eq:QS-int}
    v_\mu \equiv v_\mu(\bfr_{i,\mu},[\{\rho_\nu\}])
\end{equation}
  meaning that $v_\mu$ is generally both a function of the particle's position and a functional of all the density fields.
  Importantly, QS interactions~\eqref{eq:QS-int} are non-reciprocal in nature:
  the displacements of particles $i$ and $j$
of species $\mu$ and $\nu$ impact their respective velocities in
completely arbitrary ways. A question {that has attracted much attention recently~\cite{fodor_how_2016,nardini2017entropy,shankar2018hidden,mahdisoltani2021nonequilibrium,o2022time}} is then how this
non-reciprocity affect their large-scale properties. To address this
question, we first explicitly coarse-grain the
dynamics~\eqref{eq:microdyn} and compute the entropy production rate $\sigma$
of the resulting fluctuating hyrodynamics. We then show that this entropy
production rate \textit{vanishes} whenever
\begin{equation}\label{eq:intointegrable}
\forall (\mu,
  \nu), \qquad  \frac{\delta \log v_\mu(\bfr)}{\delta \rho_\nu(\bfr')} =
  \frac{\delta \log v_\nu(\bfr')}{\delta \rho_\mu(\bfr)}\;.
\end{equation}
In this case, the microscopic non-equilibrium model leads to an
effective large-scale equilibrium theory and non-reciprocity vanishes
upon coarse-graining. To the best of our knowledge,
condition~\eqref{eq:intointegrable} is the first non-trivial
generalization of Newton's action-reaction principle to a microscopic
model of mixtures of active particles interacting via a many-body
mediated interaction like QS. We show how the system then admits an
effective free energy which allows predicting its emerging
behavior. Remarkably, this allows constructing the phase diagram of
the system from its microscopic dynamics without any fit parameter, a
rare achievement even in equilibrium. On the contrary, when
Eq.~\eqref{eq:intointegrable} is violated, non-reciprocity survives
coarse-graining and we derive a sufficient condition on the
\textit{microscopic} dynamics to observe travelling patterns that
explicitly break time-reversal symmetry at the macroscopic scale.
}\fi


\section{Fluctuating hydrodynamics} We start by coarse-graining
the microscopic dynamics~\eqref{eq:microdyn} in the presence of QS
interactions. The hydrodynamic modes are the fluctuating conserved
density fields $\rho_\mu(\bfr)=\sum_{j}\delta(\bfr-\bfr_{j,\mu})$. As
shown in {the} Supplementary Information, the dynamics in $d$ space
dimensions can be obtained as $N$ coupled It\=o-Langevin
equations:
\begin{equation}\label{eq:diff-drift}
    \partial_t \rho_\mu = - \nabla_{\mathbf{r}} \cdot \big[  \mathbf{V}_\mu \rho_\mu - D_\mu \nabla_{\mathbf{r}} \rho_\mu + \sqrt{2 D_\mu \rho_\mu} \, \boldsymbol{\Lambda}_\mu \big]\;,
\end{equation}
where the $\boldsymbol{\Lambda}_\mu(\bfr,t)$ are independent
Gaussian white noise fields of zero mean, unit variance, and
independent components. The collective diffusivities and drifts then read
\begin{equation}\label{eq:macrotomicro}
  D_\mu = \frac{v_\mu^2(\mathbf{r}, [\{\rho_\nu\}])\tau_\mu}{d },\qquad \mathbf{V}_\mu = - D_\mu  \nabla \log[v_\mu (\mathbf{r}, [\{\rho_\nu\}])]\;.
\end{equation}
Inspection of Eq.~\eqref{eq:diff-drift} shows that it can be rewritten as
\begin{equation}\label{eq:diff-drift2}
  \partial_t \rho_\mu = \nabla_{\mathbf{r}} \cdot \big[  M_\mu \nabla \mathfrak{u}_\mu + \sqrt{2 M_\mu} \, \mathbf{\Lambda}_\mu \big]\;,
\end{equation}
where $M_\mu(\bfr)=\rho_\mu(\bfr) D_\mu(\bfr,[\{\rho_\nu\}]) $ is a
density-dependent collective mobility and $\mathfrak{u}_\mu(\bfr) =
\log[\rho_\mu(\bfr)]+\log[ v_\mu(\bfr,[\{\rho_\nu\}])]$ is an
effective chemical potential. One can then study the large-scale
behavior of the system using Eq.~\eqref{eq:diff-drift2} and connect
it to the microscopic dynamics using Eqs.~\eqref{eq:macrotomicro}. 

\section{When non-reciprocity vanishes upon coarse-graining}

Note that the fluctuating hydrodynamics~\eqref{eq:diff-drift2} takes a
form reminiscent of $N$ coupled model-B
dynamics~\cite{hohenberg_theory_1977} and it is thus natural to ask
whether we have coarse-grained our microscopic model into an effective
equilibrium one. To address this question, we show in the
Supplementary Information that the stochastic
dynamics~\eqref{eq:diff-drift2} leads to an entropy production rate
given by:
\begin{equation}\label{eq:sigmacg}
  \sigma =  \int d^d \bfr  \sum_{\mu=1}^N \Big\langle M_\mu \Big[\nabla \Big( {\mfu_\mu} + \frac{\delta \log P_s}{\delta \rho_\mu(\bfr)} \Big) \Big]^2 \> \Big\rangle\;,
\end{equation}
where $P_s[\{\rho_\nu(\bfr)\}]$ is the steady-state
distribution. Equation~\eqref{eq:sigmacg} relates the irreversibility
of the system at the macroscopic scale to its microscopic parameters
$\{v_\nu\}$ {through $\mfu_\mu = \log v_\mu + \log \rho_\mu$}. For generic
QS interactions, $\sigma$ is positive and the coarse-grained dynamics
is out of equilibrium. {However}, the model admits a macroscopic
equilibrium limit whenever there exists a functional ${\cal
  F}[\{\rho_\nu\}]$ such that $\mathfrak{u}_\mu(\bfr)=\frac{\delta
  \cal F}{\delta \rho_\mu(\bfr)}$. It is then easy to check that
$P_s\propto\exp[-{\cal F}]$ and that $\sigma$ vanishes. Furthermore,
${\cal F}$ is mathematically equivalent to a free energy in
equilibrium and plays the role of a Lyapunov functional for the
dynamics, since $\partial_t \langle{\cal F}\rangle=-\int d{\bfr}
\sum_\mu \langle M_\mu (\nabla \frac{\delta {\cal F}}{\delta
  \rho_\mu})^2\rangle <0$.

To assess whether such an effective free energy exists, we need to
determine the conditions under which $\mfu_\mu$ can be written as a
functional derivative. To do so, we generalize the functional Schwarz
theorem~\cite{o2022time} to the case of $N$ coupled stochastic field
equations. \if{To determine the conditions under which such an
  effective free energy exists, we need to derive a functional Schwarz
  theorem~\cite{o2022time} for the case of $N$ coupled stochastic
  field equations.}\fi As shown in the {Supplementary
  Information}, this leads to a system of $N^2$ equations in the sense
of distributions:
\begin{equation}\label{eq:intcond}
\forall (\mu,\nu),\qquad  \cD_{\mu\nu}(\bfr,\bfr')\equiv\frac{\delta \mfu_\mu(\bfr)}{\delta \rho_\nu(\bfr')}-\frac{\delta \mfu_\nu(\bfr')}{\delta \rho_\mu(\bfr)}=0.
\end{equation}
Using the explicit expression for the chemical potential $\mfu_\mu$
then directly leads to the condition~\eqref{eq:intointegrable} for the
microscopic dynamics. When the self-propulsion depends exclusively on
local densities, \textit{i.e.} $v_\mu(\bfr) \equiv v_\mu(\rho_1(\bfr),
\dots, \rho_N(\bfr))$, Equation~\ref{eq:intcond} simplifies {to}
\begin{equation}\label{eq:integrable}
  \frac{\partial \log v_\mu}{\partial \rho_\nu} = \frac{\partial \log
    v_\nu}{\partial \rho_\mu}\;,
\end{equation}
whose full solution space can be constructed explicitly. Indeed, the
solutions to Eq.~\eqref{eq:integrable} are generated by the gradient
in $\rho_\mu$-space of all the `potentials' $U(\rho_1,\dots,\rho_N)$
through $\log v_\mu=\frac{\partial
  U}{\partial {\rho_\mu}}$\if{$(\log v_1,\dots,\log v_N)=\nabla_{\!\boldsymbol{\rho}} U$,
where $(\nabla_{\!\boldsymbol{\rho}} U)_{\mu} \equiv \frac{\partial
  U}{\partial {\rho_\mu}}$}\fi.  The effective free energy can then be
directly computed as
\begin{subequations}\label{eq:FreeEnergyTot}
\begin{align}\label{eq:FreeEnergy}
  {\cal F}[\{\rho_\mu\}]&= \int d\bfr f(\bfr)\\
  f(\bfr)&=U(\{ \rho_\mu(\bfr)\})+\sum_\mu \rho_\mu(\bfr) \log\rho_\mu(\bfr)\;. \label{eq:FreeEnergyDensity}
\end{align}
\end{subequations}
Importantly, even though the microscopic dynamics then maps onto a
\textit{bona fide} equilibrium problem at the macroscopic scale, the
contribution of particle $(i,\mu)$ to the velocity $\dot\bfr_{j,\nu}$ of
particle $(j,\nu)$ is \textit{not} equal and opposite to the
contribution of particle $(j,\nu)$ to $\dot\bfr_{i,\mu}$.
In other
words, momentum conservation is still violated at the microscopic
scale and the interactions are still non-reciprocal.
Equation~\eqref{eq:integrable} can then be seen as a non-trivial
\textit{microscopic constraint} on the QS interactions such that
action-reaction is restored at the \textit{macroscopic scale}.

Let us now show how our effective equilibrium theory allows us to
account for the emerging behaviors of binary active mixtures when
equation~\eqref{eq:intointegrable} is satisfied. For sake of
generality, we allow both for global interactions, where the
self-propulsion speed $v_\mu$ of species $\mu$ depends on the total
density field of particles, $\rho_{\rm t}(\bfr)=\sum_\mu
\rho_\mu(\bfr)$, and for specific ones, where $v_\mu$ depends
specifically on one---or more---density {field} $\rho_\nu(\bfr)$. In the
latter case, we refer to self and cross interactions when $\mu=\nu$
and $\mu\neq \nu$, respectively. We consider self-inhibition of
motility coupled to a global enhancement of motility through
\begin{equation}\label{eq:vreciprocal}
  v_{\mu}(\bfr)=v_\mu^0 \phi_\mu^s[\tilde \rho_\mu(\bfr)] \phi_\mu^g[\tilde\rho_{\rm t}(\bfr)]\;.
\end{equation}
In Eq.~\eqref{eq:vreciprocal}, self and global regulations are
modelled by sigmoidal functions $\phi_\mu^{s}$ and
$\phi_\mu^{g}$, respectively, and $\tilde \rho(\bfr) = K
\ast \rho(\bfr)$ is a local measurement of the density field obtained
by convolution with a kernel $K$ (see Supplementary Information for details). Note that
QS interactions leading to Eq.~\eqref{eq:vreciprocal} can actually be
{realized} using orthogonal QS circuits starting from clonal bacterial
strains~\cite{curatolo_cooperative_2020}. The non-local sampling of
the density then results from a fast variable treatment on the
signalling molecular field~\cite{obyrne_lamellar_2020}. Alternatively,
such interactions can be directly engineered for light-controlled
active
colloids~\cite{bauerle_self-organization_2018,lavergne_group_2019}.
Self-inhibition and global activation of motility then correspond to
$\partial_{\tilde\rho}\phi_\mu^s<0$ and $\partial_{\tilde
  \rho}\phi_\mu^g>0$, respectively. To map out the phases accessible
to the system, we carried out large-scale simulations of
dynamics~\eqref{eq:microdyn} using the QS
interactions~\eqref{eq:vreciprocal}, as described in Supplementary Information.

\begin{figure}
  \begin{center}
    \begin{tikzpicture}
      \def\x{2.7}
      \def\start{-0.4}
      \path (\start,0) node {\includegraphics[width=.33\columnwidth]{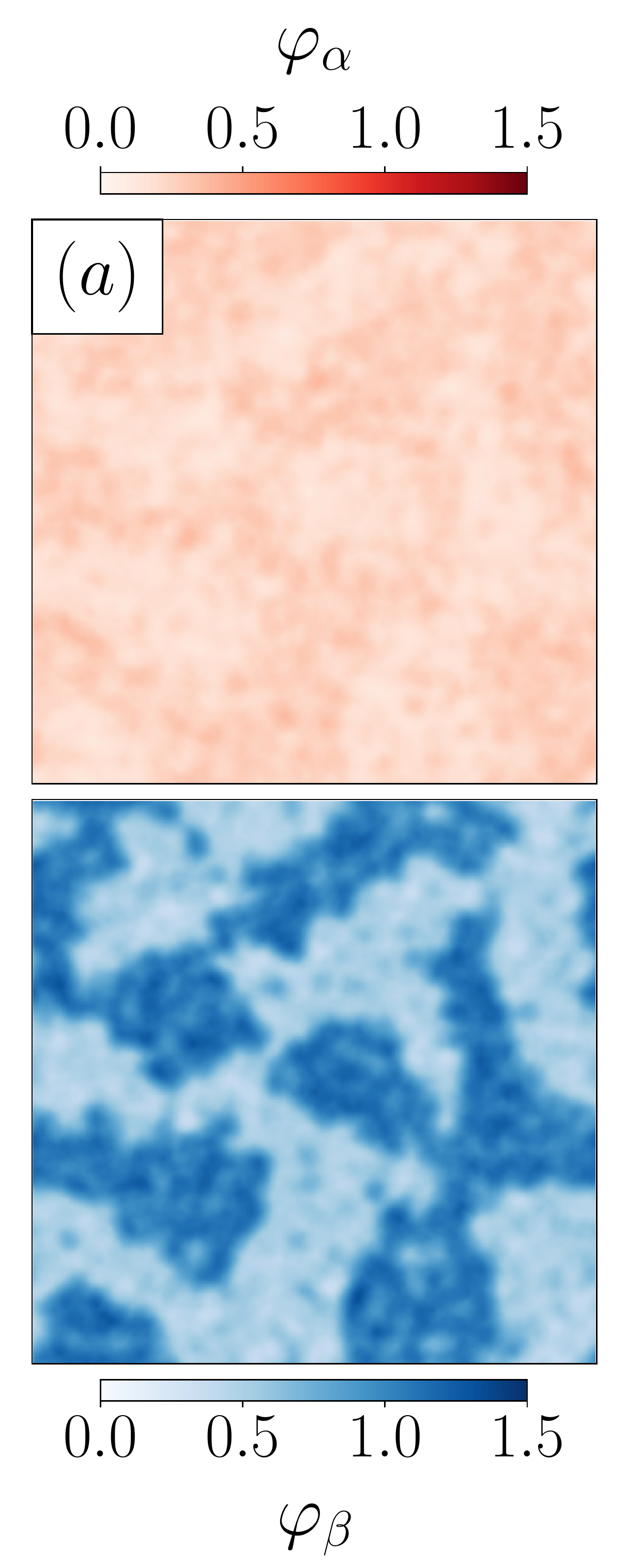}};
      \path (\start+\x,0) node {\includegraphics[width=.33\columnwidth]{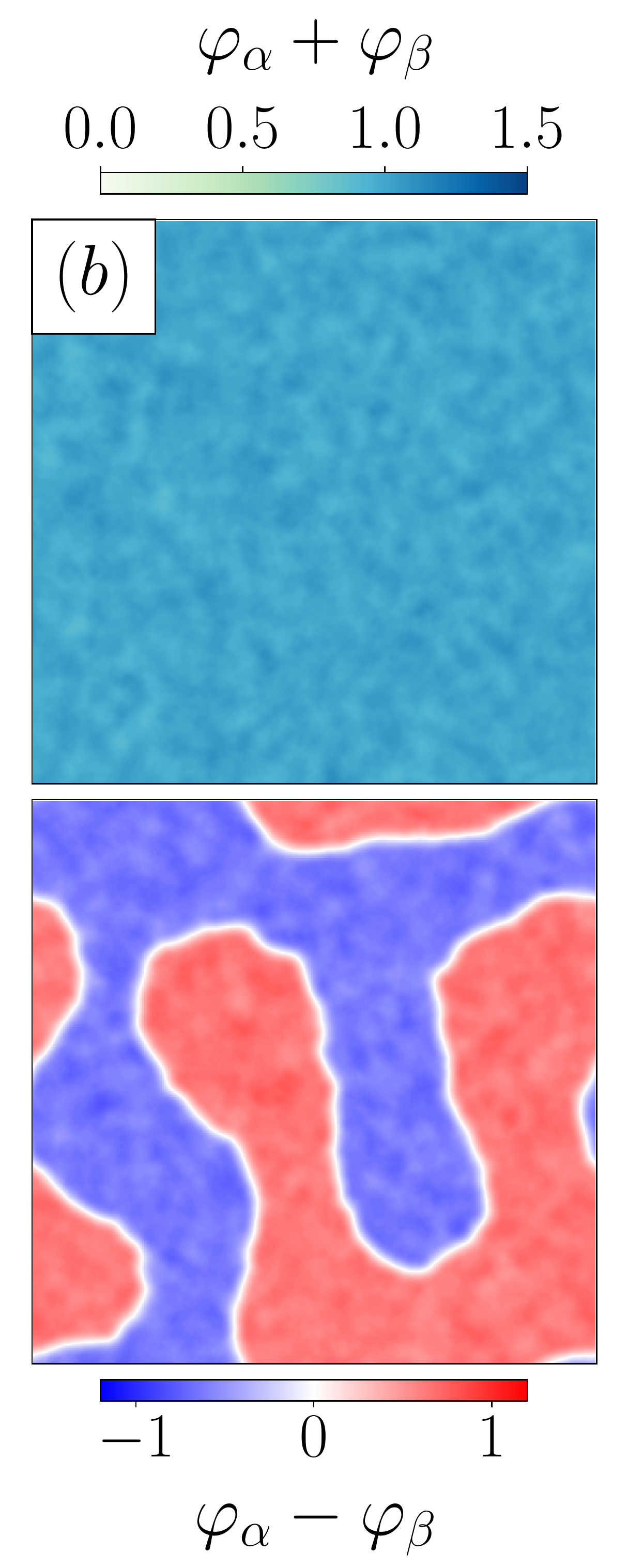}};
      \path (\start+2*\x,0) node {\includegraphics[width=.33\columnwidth]{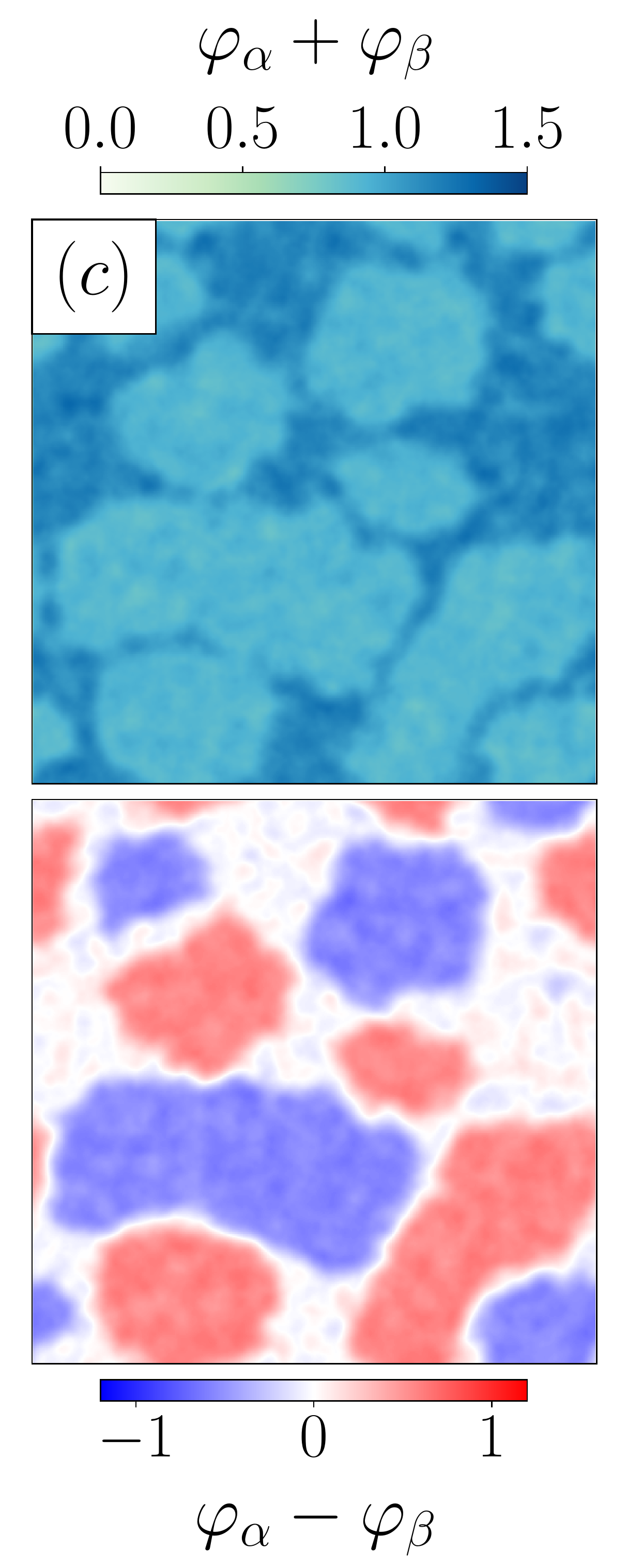}};
    \end{tikzpicture}
  \end{center}
  \caption{Simulations of two species of RTPs with self-inhibition and
    global activation of the motility as in
    Eq.~\eqref{eq:vreciprocal}, for different average densities
    $\rho_{\alpha,\beta}^0$. {Normalized densities are defined as:
      $\varphi_{\mu}=\rho_\mu/(\rho_\alpha^0+\rho_\beta^0)$.} \textbf{(a)}
    Phase separation of species $\beta$ ($\rho_\alpha^0=15$,
    $\rho_\beta^0=50$). \textbf{(b)} Demixing
    ($\rho_{\alpha,\beta}^0=55$). \textbf{(c)} Triple coexistence between
    $\alpha$-rich, $\beta$-rich, and well-mixed phases
    ($\rho_{\alpha,\beta}^0=75$).  All parameters and numerical
    details are given in Supplementary Information.}\label{fig:phenomenology}
\end{figure}

Varying the overall composition of the system reveals a rich
phenomenology. First, as in single-species systems, the
self-inhibition of a given species may lead to its motility-induced
phase separation~\cite{cates_motility-induced_2015,fodor2018statistical}. The
second species then experiences a mild opposite modulation of its
density field (Fig.~\ref{fig:phenomenology}a). Then, two phases
specific to (active) mixtures emerge from the global coupling. First,
Fig.~\ref{fig:phenomenology}b shows a segregated phase in which the
two strains demix and undergo phase separation. Note that the global
density field $\rho_{\rm t}(\bfr)$ remains homogeneous, contrary to
what happens in the case of single-species phase separation shown in
Figs~\ref{fig:phenomenology}a and S1. Second,
Fig.~\ref{fig:phenomenology}c shows the existence of a
triple-coexistence regime leading to the joint observation of
$\alpha$-rich, $\beta$-rich, and well-mixed phases, together with an
overall phase-separation for $\rho_t(\bfr)$. Let us now show how our
effective equilibrium theory allows us to account for the phase
diagram of the system quantitatively.


Under the local approximation $\tilde\rho_\mu(\bfr)\simeq
\rho_\mu(\bfr)$, Eq.~\eqref{eq:vreciprocal} leads to $\log v_\mu=\log
v_\mu^0 + \log \phi_\mu^s(\rho_\mu)+\log\phi_\mu^g(\rho_{\rm
  t})$. Direct inspection shows that Eq.~\eqref{eq:integrable} then
amounts to $\phi_\mu^g=\phi_\nu^g$ for all $\mu,\nu$: the sole
requirement for effective equilibrium is that the global interaction
term $\phi_\mu^g$ be common to all species. This is the case for the
system shown in Fig.~\ref{fig:phenomenology}a-c, which can thus be
mapped onto an equilibrium problem.  The self-organization of the two
coexisting species can then be predicted from the analysis of the
corresponding effective free energy, which we detail in the
Supplementary Information. Departure from a homogeneous, well-mixed system {will occur}
whenever the free energy density~\eqref{eq:FreeEnergyDensity} is not
convex. Predicting the coexisting densities then amounts to
constructing the tangent planes of
$f(\rho_\alpha,\rho_\beta)$~\cite{sollich_predicting_2001}, as
detailed in the Supplementary Information. In
Fig.~\ref{fig:equil_phase_diagram}a, we compare these theoretical
predictions to direct measurements for the parameters corresponding to
Fig.~\ref{fig:phenomenology}a-c. Despite the construction relying on a
locality asumption and a mean-field approximation, the agreement
between predicted and measured phase diagrams in the composition space
$(\rho_\alpha^0,\rho_\beta^0)$ is excellent. The triple coexistence
regime reported in Fig.~\ref{fig:phenomenology}c emerges when the free
energy surface admits a plane that is tangent in three points
$(\rho_\alpha^i,\rho_\beta^i)_{i\in \{1,2,3\}}$ simultaneously, as
illustrated in Fig.~\ref{fig:equil_phase_diagram}b. The corresponding
compositions then delimit the coexistence region and determine the
coexisting phases, while the respective fraction of each phase is 
obtained using the lever rule. Finally, the existence of an effective
free energy also ensures that the Gibbs phase {rule} applies, which explains
the existence of the three-phase and two-phase coexistence regions for our
active binary mixture. The equilibrium mapping thus fully accounts for
the static phase-separation scenario reported in
Fig.~\ref{fig:phenomenology}. We now illustrate how violations of the
microscopic condition~\eqref{eq:integrable} may lead to an emerging
physics that explicitly breaks time-reversal symmetry.

\begin{figure}
  \begin{center}
    \begin{tikzpicture}
      \def\x{4.5}
      \path (0,0) node {\includegraphics[width=.52\columnwidth]{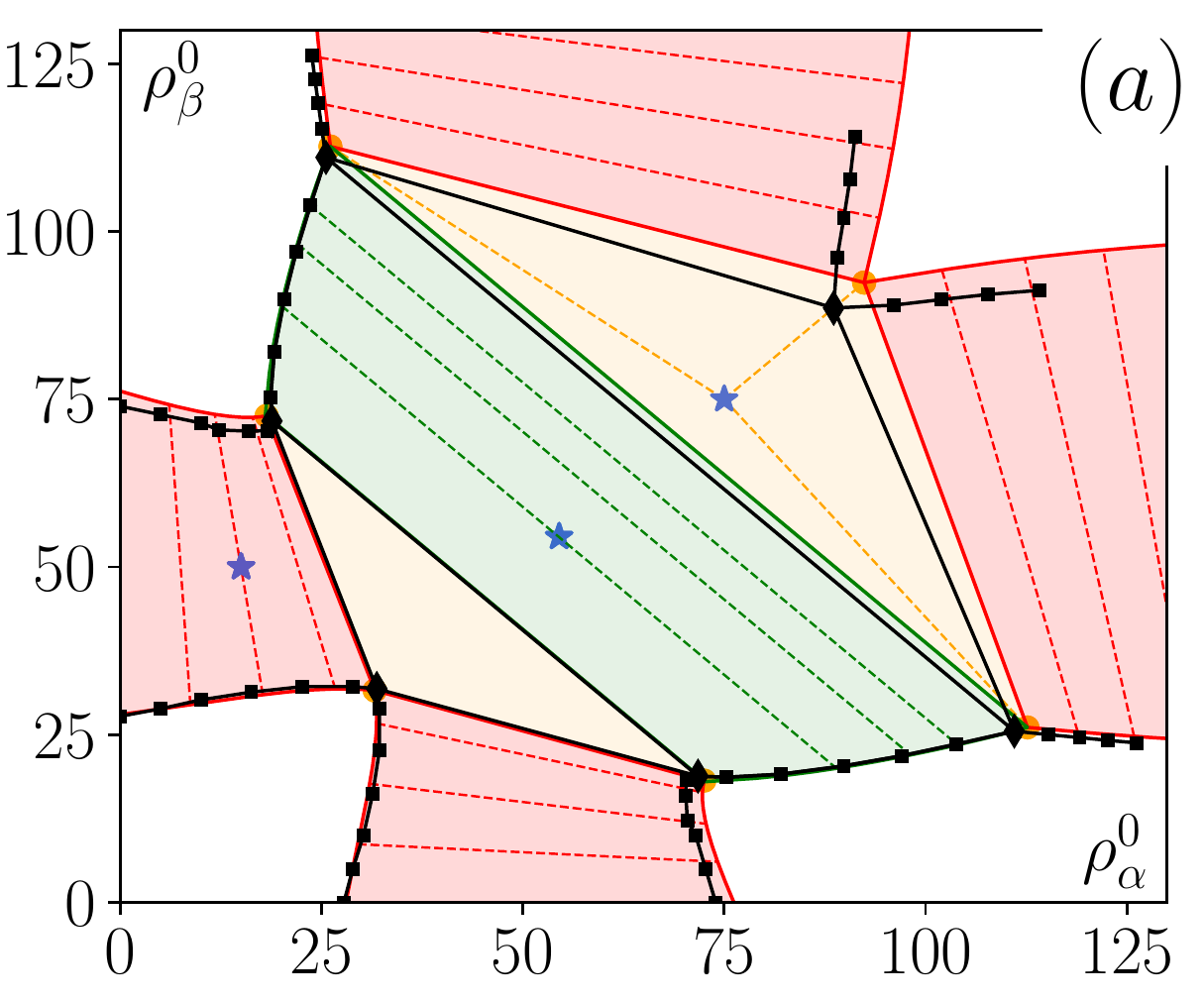}};
      \path (\x,0) node {\includegraphics[width=.52\columnwidth]{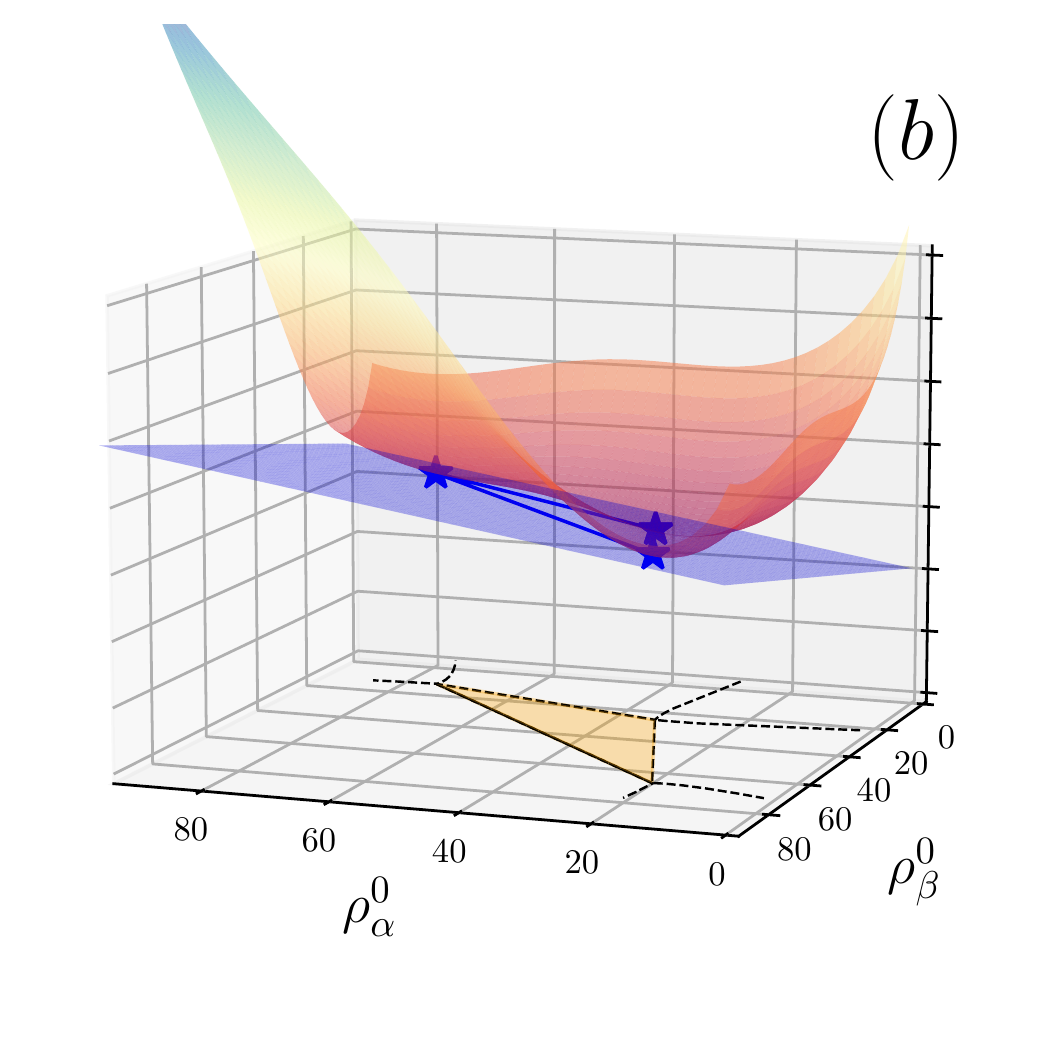}};
    \end{tikzpicture}
  \end{center}
  \caption{{\bf (a)} Phase diagram of two species of RTPs experiencing
    self-inhibition and global activation of motility according to
    Eq.~\eqref{eq:vreciprocal}. White regions correspond to
    homogeneous well mixed phases. Red, green, and ochre regions
    {indicate} one-species phase separation, demixing, and triple
    phase coexistence, respectively. Stars correspond to snapshots
    shown in Fig.~\ref{fig:phenomenology}a-c. Coexistence lines {(solid)}
    and tie-lines {(dashed)} are predicted using a tangent plane
    construction on the free energy density
    $f(\rho_\alpha,\rho_\beta)$ as detailed in {the} Supplementary
    Information. Black squares {show} coexisting densities
    measured in simulations. {\bf (b)} Plot of the free energy density
    in the triple coexistence regime from
    Fig.~\ref{fig:phenomenology}c. The points where the tangent plane
    in blue touches the surface determine the three compositions that
    will be observed in the coexistence region.}
  \label{fig:equil_phase_diagram}
\end{figure}

\section{When non-reciprocal interactions survive coarsening}

The violation of Eq.~\eqref{eq:intointegrable} is a sufficient
condition for the emergence of non-reciprocal couplings between the
density fields at the macroscopic
scale~\eqref{eq:diff-drift2}. Consequently, the entropy production
rate~\eqref{eq:sigmacg} is positive. The lack of a gradient structure
for hydrodynamic equations is well known to allow for the existence of
travelling
patterns~\cite{cross_pattern_1993,saarloos_amplitude_1994,aranson2002world,saha_scalar_2020,you_nonreciprocity_2020,frohoff2021suppression,frohoff2021localized,fruchart_non-rec_2021}. To
determine the \textit{microscopic} condition for these to emerge, we
consider the fate of perturbations around homogeneous solutions of
Eq.~\eqref{eq:diff-drift2},
$\rho_\mu(\bfr)=\rho_\mu^0+\delta\rho_\mu(\bfr)$. In Fourier space,
their linearized dynamics read $\partial_t \delta\hat\rho_\mu(\bfq)=
-\bfq^2 \mathcal{M}_{\mu\nu}(\bfq) \delta\hat\rho_\nu$, with
\begin{equation}\label{eq:LSupplementary Information}
\mathcal{M}_{\mu\nu}(\bfq)=D_\mu^0\Big[\delta_{\mu\nu}+ \rho_\mu^0
\frac{\partial}{\partial \rho_\nu} \log v_\mu\Big]\;\,
\end{equation}
where $D_\mu^0\equiv D_\mu(\{\rho_\nu^0\})$. As shown in Supplementary Information, for
$N=2$ species, eigenvalues with non-vanishing imaginary parts
require
\begin{equation}\label{eq:LScomplex}
  \frac{\partial v_\beta }{\partial \rho_\alpha} \frac{\partial v_\alpha}{\partial \rho_\beta} <-\frac{v_\alpha v_\beta (\mathcal M_{\alpha\alpha}-\mathcal M_{\beta\beta})^2}{4 D_\alpha^0 D_\beta^0 \rho_\alpha^0 \rho_\beta^0}\;.
\end{equation}
We thus predict an oscillatory behavior in the presence of sufficiently strong, opposite interactions, e.g. when species $\beta$ enhances the speed
of $\alpha$ while $\alpha$ inhibits the motility of $\beta$.
Under
this condition, which is a stronger requirement than simple non-reciprocity, homogeneous profiles are linearly unstable whenever
\begin{equation}\label{eq:LSinstab}
  \sum_{\mu=\alpha,\beta} D_\mu^0 (1 + \rho_\mu^0
\frac{\partial}{\partial \rho_\mu} \log v_\mu) <0\;.
\end{equation}
This opens up the possibility of travelling patterns, which we now
explore.

\onecolumngrid

\begin{figure}
  \begin{tikzpicture}
    \def\x{6}
    \path (0,0)  node {\includegraphics[width=.32\textwidth]{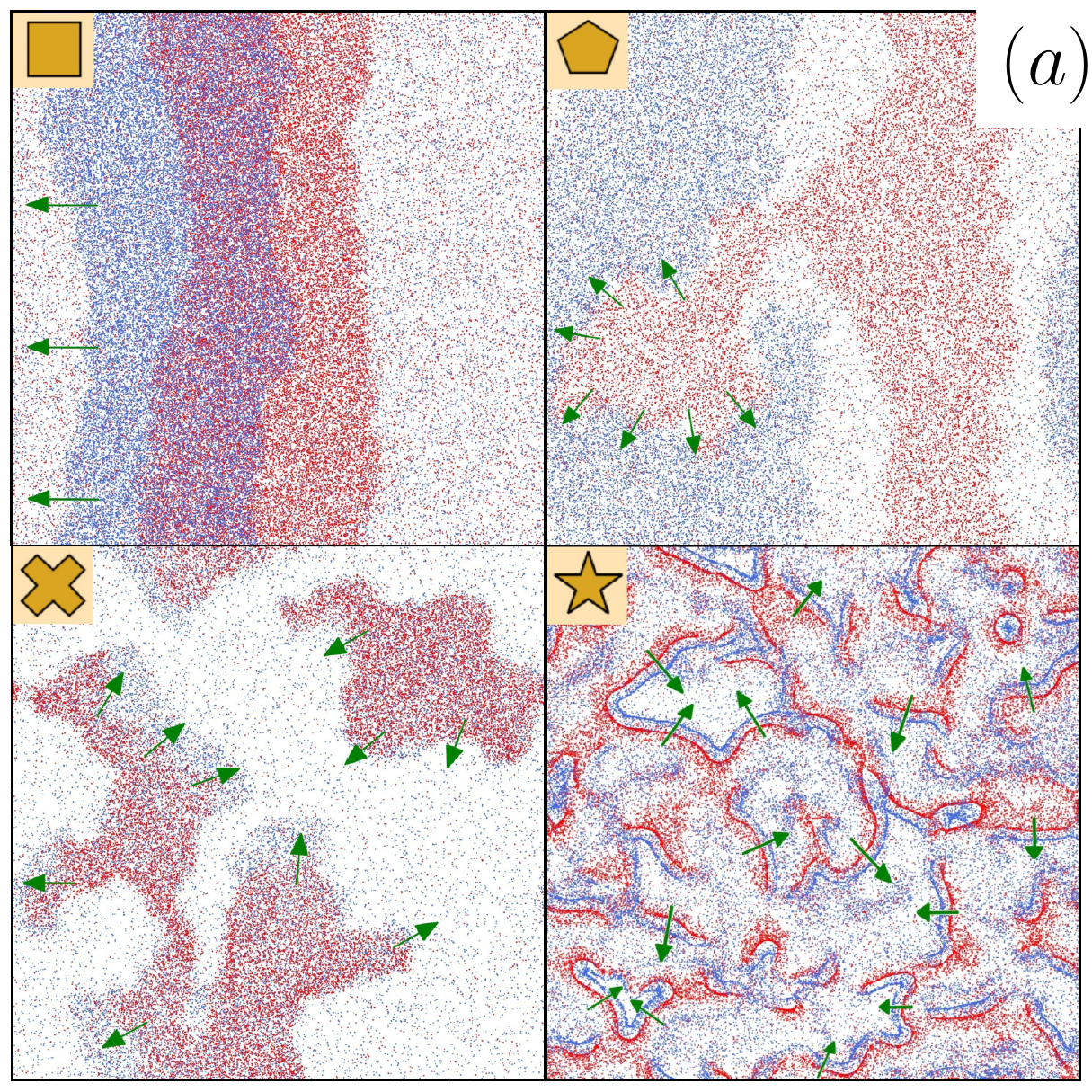}};
    \path (\x,0) node {\includegraphics[width=.32\textwidth]{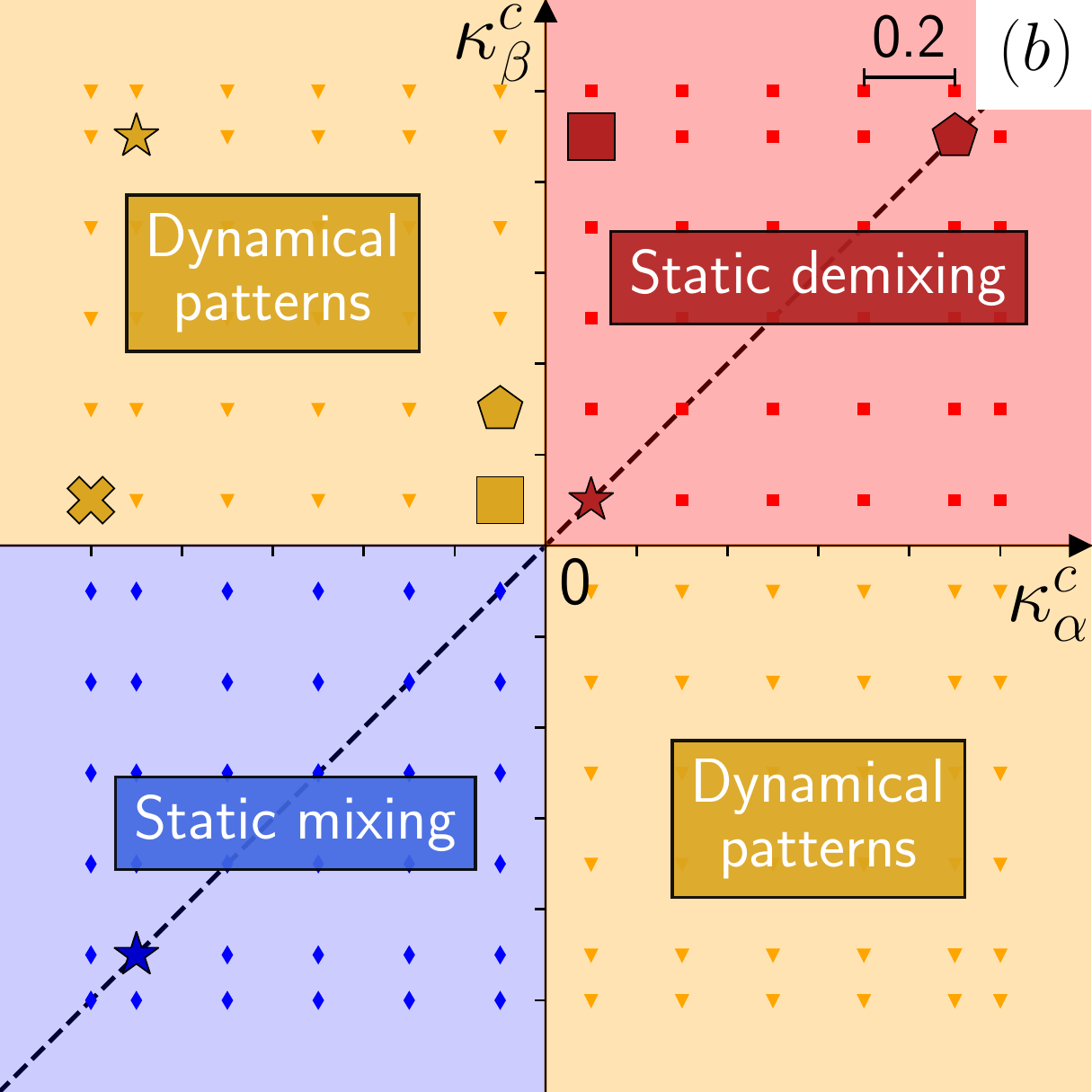}};
    \path (2*\x,0) node {\includegraphics[width=.32\textwidth]{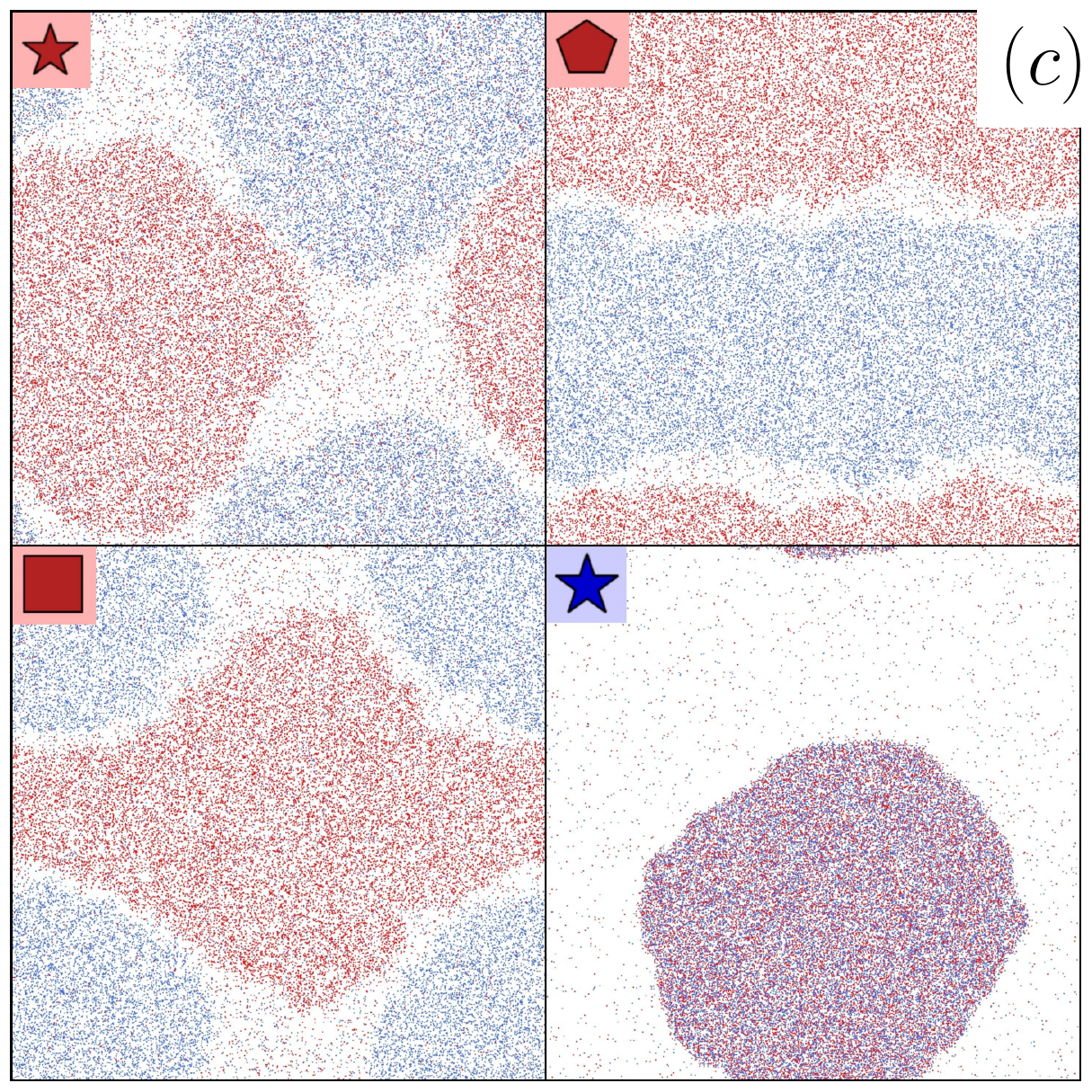}};
  \end{tikzpicture}
  \caption{Microscopic simulations of Eq.~\eqref{eq:vcross}, with
    self-inhibition and non-reciprocal cross interactions. Panels {\bf
      (a)} and {\bf (c)} show the dynamical and static patterns
    observed in simulations, respectively; $\alpha$-particles are
    depicted in red, $\beta$-particles in blue. The snapshots
    correspond to the larger symbols shown in panel (b).  {\bf (b)}
    Phase diagram as the couplings $\kappa_\alpha^c$ and
    $\kappa_\beta^c$ are varied, where we remind that $\kappa_\mu^c>0$ corresponds to the activation of the motility of species $\mu$ by the density of species $\nu$, whereas $\kappa_\mu^c<0$ corresponds to an inhibition. The background colors correspond to
    the prediction of linear stability analysis which are confirmed by
    numerical simulations (small symbols).  The phase diagram is
    symmetric with respect to the dashed line
    $\kappa_\beta^c=\kappa_\alpha^c$ upon inverting the roles of
    $\alpha$- and $\beta$-particles in panel (a,c).  See Supplementary Information for
    other parameters and numerical details.}
  \label{fig:travelling}
\end{figure}
\twocolumngrid

To do so, we carry out simulations of a two-species system with
self-inhibition of motility and non-reciprocal cross-interactions
given by
\begin{equation}\label{eq:vcross}
v_{\mu}(\bfr)=v^0_\mu \phi_\mu^s[\tilde \rho_\mu(\bfr)] \phi^c_\mu[\tilde \rho_{\nu}(\bfr)]\;.
\end{equation}
To control the strength of the non-reciprocal couplings, we chose
$\phi^c_\mu(\tilde\rho)=\exp[\kappa^c_\mu {\cal
    S}_\mu^c(\tilde\rho)]$, where ${\cal S}_\mu^c(\tilde\rho)$ is a
sigmoidal function described in Supplementary Information, and we vary the values of
$\kappa^c_\mu$.  We use self-inhibitions strong enough for 
Eq.~\eqref{eq:LSinstab} to hold so that the system is never
homogeneous. As show in Figure~\ref{fig:travelling}, our simulations
reveal a variety of static and dynamical patterns. In agreement with
our prediction~\eqref{eq:LScomplex}, travelling patterns are observed
when $\kappa_\alpha^c\kappa_\beta^c<0$ (orange quadrants). An
intuitive microscopic understanding of the observed phases can be
achieved by noticing that each species tends to accumulate where it
goes
more slowly~\cite{schnitzer_theory_1993,tailleur_statistical_2008,cates_when_2013,martin_PRE_2021,frangipane_dynamic_2018,arlt_painting_2018}. Motility
inhibition then acts as an effective attraction whereas motility
enhancement leads to effective repulsion. When
$\kappa_{\alpha,\beta}^c$ are both positive, the species effectively
repel each other, leading to a triple coexistence regime with demixing
between the dense phases (red quadrant). On the contrary, negative
$\kappa_{\alpha,\beta}^c$ lead to effective attractive interactions
and colocalization of the liquid phases (blue quadrant). The
frustrated case, $\kappa_\alpha^c\kappa_\beta^c<0$, 
corresponds to the motility of one species---say $\alpha $---being inhibited by the
other---$\beta$ in this case---while {that of} $\beta$ is enhanced by
$\alpha$. This leads to a complex run-and-chase dynamics between the
two species that results in steady (orange square) or chaotic (orange
star) travelling bands when $|\kappa_\alpha^c|\simeq|\kappa_\beta^c|$,
as well as to a {rich} variety of more complex dynamical behaviors in
less symmetric cases (orange cross and pentagon as well as SI movie
3-5). Thanks to the explicit coarse-graining of the microscopic
dynamics, we are thus able to determine the microscopic condition for
travelling patterns to emerge and to identify the mechanism leading to
the run-and-chase dynamics.

\if{Finally, while the presence of self-inhibition allows one
to rationalize the emergent behaviors reported in
Figure~\ref{fig:phenomenology}d and~\ref{fig:travelling}, non-linear
dynamical patterns can also arise when homogeneous profiles are linear
stable~\cite{supp}. Note that there are strong signs that the
  instability leading to band chaos is always present for
  $\kappa^c\neq 0$, as can be checked by systematically increasing
  system size starting from the band reported in
  Fig.~\ref{fig:phenomenology}d, a scenario reminiscent of what is
  observed in active nematics~\cite{ngo2014large}.}\fi

\section{Non-reciprocity in chemotaxis.}
To show how our results generalize beyond the case of quorum sensing,
we consider chemotactic interactions which have attracted a lot of
interest in the context of bacterial
suspensions~\cite{berg1975chemotaxis,budrene1995dynamics,woodward1995spatio,brenner1998physical,saragosti2011directional,chatterjee2011chemotaxis,sourjik2012responding,cremer2019chemotaxis}. For
sake of generality, we consider $N$ {species of} ABPs/RTPs that evolve according to {the}
dynamics~\eqref{eq:microdyn} and whose motilities are biased by the
gradients of $n$ chemical fields $\{c_p(\bfr)\}$. We allow both for
biases on the reorientation dynamics and on the self-propulsion
speeds:
\begin{equation}
  \begin{split}
    v_\mu =& v_{0\mu} - \bfu_{i,\mu} \cdot \sum_{p=1}^n v_{1\mu}^p
    \nabla_{\bfr_{i,\mu}} c_p \; ,\\ \tau^{-1}_\mu =&
    \tau^{-1}_{0\mu} + \bfu_{i,\mu} \cdot \sum_{p=1}^n
    (\tau_{1\mu}^p)^{-1} \nabla_{\bfr_{i,\mu}} c_p \>,
    \label{eq:CT-interactions}
    \end{split}
  \end{equation}
  where the parameters $v_{*\mu}$, $\tau_{*\mu}^{-1}$ are constant.
  When $v_{1\mu}^p$, $\tau_{1\mu}^p$ are positive, particles increase
  their persistence lengths when moving towards minima of $c_p$,
  implying that $c_p$ acts as a chemorepellent. Conversely, negative
  values of $v_{1\mu}^p$, $\tau_{1\mu}^p$ correspond to
  chemoattraction. We consider the case in which the chemicals are
  produced by the particles before they diffuse and degrade in the
  environment at non-vanishing rates. In the large
  system-size {limit}, the dynamics of $c_p$ is thus much faster than
  that of the conserved density field $\rho_\mu$ and the chemical
  fields follow the evolution of the density fields {adiabatically}:
  $c_p(\bfr) \equiv c_p(\bfr,[\{\rho_\nu\}])$.

  \if{
  Typically, $c$ represents
  the concentration of a chemical signal that is produced or consumed
  by the particles themselves.  It is thus reasonable to assume that
  the chemical dynamics occurs over a time-scale which is much faster
  than the particle dynamics\cite{obyrne_lamellar_2020}, leading to $c
  \equiv c(\bfr_{i,\mu},[\{\rho_\nu\}])$ in
  Eq~\eqref{eq:CT-interactions}.}\fi

  We start by coarse-graining the microscopic dynamics into a
  stochastic field theory for the density fields, which takes the form
  of Eq.~\eqref{eq:diff-drift2} with an effective chemical potential
  given by (see Supplementary Information):
  \begin{equation}\label{eq:chempotchemot}
    \mathfrak{u}_\mu = \frac{1}{v_{0\mu}^{\>\>2}} \sum_{p=1}^n
    \left(\frac{v_{1\mu}^p}{\tau_{0\mu}} +
    \frac{v_{0\mu}}{\tau_{1\mu}^p} \right) c_p + \log \rho_\mu \;.
  \end{equation}
  Consequently, the entropy production rate remains given by
  Eq.~\eqref{eq:sigmacg} albeit with $\mathfrak{u}$ {determined} by
  Eq.~\eqref{eq:chempotchemot}. The integrability condition for
  non-reciprocity to vanish across scales is then given by the
  functional Schwarz theorem as:
  \begin{equation}\label{eq:intointegrable-CT}
    \begin{split}
 \forall (\mu,
  \nu), \qquad  \frac{1}{v_{0 \mu}^2} \sum_{p=1}^n \left(\frac{v_{1\mu}^p}{\tau_{0\mu}} +
  \frac{v_{0\mu}}{\tau_{1\mu}^p} \right) \> \frac{\delta c_p(\bfr)}{\delta \rho_\nu(\bfr')} =\\
  \frac{1}{v_{0\nu}^2}  \sum_{p=1}^n \left(\frac{v_{1\nu}^p}{\tau_{0\nu}} +
  \frac{v_{0\nu}}{\tau_{1\nu}^p} \right) \> \frac{\delta c_p(\bfr')}{\delta \rho_\mu(\bfr)}\;.
  \end{split}
\end{equation}
Equation~\eqref{eq:intointegrable-CT} determines when a
microscopic chemotactic dynamics admits a
large-scale effective equilibrium description.

\if{To better understand this,
  let us focus on the left-hand side of
  Eq.~\eqref{eq:intointegrable-CT}.  The functional derivative $\delta
  c/\delta \rho_\nu$ physically tells us how the chemical field $c$
  responds to fluctuations in $\rho_\nu$. The prefactor, insted,
  captures the chemotactic response of a $\mu$ particle to the
  $c$-gradients. On the whole, the left-hand side of
  Eq.~\eqref{eq:intointegrable-CT} can be interpreted as the way
  $\mu$-particles respond to fluctuations in the density of $\nu$;
  similarly, the right-hand side represents the response of
  $\nu$-particles to fluctuations in
  $\rho_\mu$. Eq.~\eqref{eq:intointegrable-CT} can thus be interpreted
  as an action-reaction statement at the field level for inter-species
  chemotactic interactions.}\fi

For sake of concreteness, let us consider the simplest case of a
single chemical field ($n=1$) given by:
\begin{equation}\label{eq:local-c}
  c(\bfr,[\{\rho_\nu\}]) = \sum_\mu \beta_\mu \tilde{\rho}_\mu(\bfr)\;,
\end{equation}
where $\tilde{\rho}_\mu = K \ast \rho_\mu$, $\beta_\mu$ is the
production rate of $c$ by species $\mu$, and $K$ is the Green's
function corresponding to the linear transport and degradation of the
chemicals. For the sake of simplicity, we only consider  biases on the
self-propulsion speeds and set $v_{0\mu} \equiv v_0$, $\tau_{0\mu}
\equiv \tau_0$ and $\tau_{1\mu}^{-1}=0$ for all species.  In the particle
dynamics, chemotactic interactions can then be seen as ``generalized''
pairwise forces:
\begin{equation}
  \begin{split}
    \dot{\bfr}_{i,\mu} =& v_0 \bfu_{i,\mu} + \sum_{j,\nu} \mathbf{f}_{\> i,\mu}^{\> j,\nu}\;,\\ \text{where}& \quad \mathbf{f}_{\> i,\mu}^{\> j,\nu} = v_{1\mu} \beta_\nu \> \bfu_{i,\mu} \> [\bfu_{i,\mu} \cdot \nabla_{\bfr_{i,\mu}} K(\bfr_{i,\mu} - \bfr_{j,\nu})]\;.
  \label{eq:microdyn-CT}
    \end{split}
\end{equation}
We stress that $\mathbf{f}_{i,\mu}^{j,\nu}$ and $\mathbf{f}_{j,\nu}^{i,\mu}$ are
not collinear, since $\mathbf{f}_{\> i,\mu}^{\> j,\nu}$ is directed
along $\bfu_{i,\mu}$. The dynamics thus explicitly violates
Newton's third law at the microscopic scale. Nevertheless,
Eq.~\eqref{eq:intointegrable-CT} ensures that reciprocity is restored
at the coarse-grained scale whenever
  $v_{1\mu} \beta_\nu = v_{1\nu} \beta_\mu$ for all species.
\if{The term $v_{1\mu} \beta_\nu$ can be interpreted as the way
  species $\nu$ effectively modulates the motility of species $\mu$:
  particles of type $\mu$ respond to fluctuations of $c$ according to
  their specific chemotactic response $v_{1\mu}^{-1}$; these
  fluctuations are, in turn, caused by any other species $\nu$
  producing or consuming $c$ at rate
  $\beta_\nu$. Eq.~\eqref{eq:integrable-local-CT} can thus be read as
  a symmetry in motility regulation between each couple of
  species.}\fi We note that equilibrium limits of
chemotactic~\cite{newman2004many,chavanis2007exact} or
diffusiophoretic
dynamics~\cite{soto_self-assembly_2014,agudo-canalejo_active_2019,ouazan-reboul_non-equilibrium_2021}
have attracted a long-standing interest in the literature. Previous
results, however, relied on microscopic Langevin dynamics in which
chemotactic interactions enter directly as effective pairwise
\textit{collinear} forces, $\mathbf{f}_i^j\propto \nabla_i
G(\bfr_i-\bfr_j)$ for some function $G$. The existence of macroscopic
equilibrium limits then relies on imposing Newton's third law at the
microscopic scale. On the contrary, Eq.~\eqref{eq:intointegrable-CT}
is, to the best of our knowledge, the first condition for chemotactic
mixtures to recover reciprocity at the macroscopic scale, despite
being non-reciprocal at the microscopic one.

\section{Discussion.}

In this article we have shown how microscopic and macroscopic scales
can be quantitatively bridged for a large class of active mixtures in
the presence of mediated {non-}reciprocal interactions. This revealed a
subtle and important property of non-reciprocity: it varies strongly
across scales. {Based on this insight we derived} non-trivial {conditions on} the {{\em microscopic}} {NRI that lead to} effective equilibrium at the {{\em macroscopic}}
scale. {This allowed} us to account---accurately and without {fit}
parameters---for the {full range} of static patterns observed in our
simulations. Finally, we derived conditions for NRI to survive
coarse-graining, hence leading to positive entropy production rate at
the macroscopic scale. When non-reciprocity is strong enough, we showed
the emergence of a wealth of dynamical patterns. Again, our
micro-to-macro approach allows us to predict the phase diagram from
microscopics without fitting parameters.

From a biophysical perspective, our study shows how QS and chemotactic
interactions lead to a rich phenomenology in complex assemblies of
cells.  In the context of bacterial colonies, motility-induced
patterns will eventually interact with population
dynamics~\cite{cates2010arrested,liu_sequential_2011,curatolo_cooperative_2020}
and genetics~\cite{hallatschek_genetic_2007}. How this interplay will
result in diverse co-existing communities is a fascinating research
direction for the future indeed.

Finally, turning synthetic active-matter systems into smart materials
will require quantitative control over complex assemblies of active
constituents. Our work demonstrates that one can go up the complexity
ladder while retaining an analytical framework to account for the
emerging properties of active systems. How these systems can then be
optimized to accomplish given tasks is an exciting challenge {that appears within reach,  given recent progress} in automatic
differentiation~\cite{goodrich2021designing}.


\textit{Acknowledgment:} JT acknowledges the financial support of ANR
Thema. AD acknowledges an international fellowship from Idex Universite
de Paris. PS acknowledges support by a RSE Saltire Facilitation Network Award. YZ acknowledges support from start-up grant NH10800621 from Soochow University.

\bibliographystyle{naturemag}       
\bibliography{biblio_draft_active_mixtures2.bib}

\begin{thebibliography}{10}
\expandafter\ifx\csname url\endcsname\relax
  \def\url#1{\texttt{#1}}\fi
\expandafter\ifx\csname urlprefix\endcsname\relax\def\urlprefix{URL }\fi
\providecommand{\bibinfo}[2]{#2}
\providecommand{\eprint}[2][]{\url{#2}}

\bibitem{o2022time}
\bibinfo{author}{O’Byrne, J.}, \bibinfo{author}{Kafri, Y.},
  \bibinfo{author}{Tailleur, J.} \& \bibinfo{author}{van Wijland, F.}
\newblock \bibinfo{title}{Time irreversibility in active matter, from micro to
  macro}.
\newblock \emph{\bibinfo{journal}{Nature Reviews Physics}}
  \textbf{\bibinfo{volume}{4}}, \bibinfo{pages}{167--183}
  (\bibinfo{year}{2022}).

\bibitem{deseigne2010collective}
\bibinfo{author}{Deseigne, J.}, \bibinfo{author}{Dauchot, O.} \&
  \bibinfo{author}{Chat{\'e}, H.}
\newblock \bibinfo{title}{Collective motion of vibrated polar disks}.
\newblock \emph{\bibinfo{journal}{Physical review letters}}
  \textbf{\bibinfo{volume}{105}}, \bibinfo{pages}{098001}
  (\bibinfo{year}{2010}).

\bibitem{schaller2010polar}
\bibinfo{author}{Schaller, V.}, \bibinfo{author}{Weber, C.},
  \bibinfo{author}{Semmrich, C.}, \bibinfo{author}{Frey, E.} \&
  \bibinfo{author}{Bausch, A.~R.}
\newblock \bibinfo{title}{Polar patterns of driven filaments}.
\newblock \emph{\bibinfo{journal}{Nature}} \textbf{\bibinfo{volume}{467}},
  \bibinfo{pages}{73--77} (\bibinfo{year}{2010}).

\bibitem{sumino2012large}
\bibinfo{author}{Sumino, Y.} \emph{et~al.}
\newblock \bibinfo{title}{Large-scale vortex lattice emerging from collectively
  moving microtubules}.
\newblock \emph{\bibinfo{journal}{Nature}} \textbf{\bibinfo{volume}{483}},
  \bibinfo{pages}{448--452} (\bibinfo{year}{2012}).

\bibitem{bricard_emergence_2013}
\bibinfo{author}{Bricard, A.}, \bibinfo{author}{Caussin, J.-B.},
  \bibinfo{author}{Desreumaux, N.}, \bibinfo{author}{Dauchot, O.} \&
  \bibinfo{author}{Bartolo, D.}
\newblock \bibinfo{title}{Emergence of macroscopic directed motion in
  populations of motile colloids}.
\newblock \emph{\bibinfo{journal}{Nature}} \textbf{\bibinfo{volume}{503}},
  \bibinfo{pages}{95--98} (\bibinfo{year}{2013}).

\bibitem{theurkauff2012dynamic}
\bibinfo{author}{Theurkauff, I.}, \bibinfo{author}{Cottin-Bizonne, C.},
  \bibinfo{author}{Palacci, J.}, \bibinfo{author}{Ybert, C.} \&
  \bibinfo{author}{Bocquet, L.}
\newblock \bibinfo{title}{Dynamic clustering in active colloidal suspensions
  with chemical signaling}.
\newblock \emph{\bibinfo{journal}{Physical review letters}}
  \textbf{\bibinfo{volume}{108}}, \bibinfo{pages}{268303}
  (\bibinfo{year}{2012}).

\bibitem{palacci2013living}
\bibinfo{author}{Palacci, J.}, \bibinfo{author}{Sacanna, S.},
  \bibinfo{author}{Steinberg, A.~P.}, \bibinfo{author}{Pine, D.~J.} \&
  \bibinfo{author}{Chaikin, P.~M.}
\newblock \bibinfo{title}{Living crystals of light-activated colloidal
  surfers}.
\newblock \emph{\bibinfo{journal}{Science}} \textbf{\bibinfo{volume}{339}},
  \bibinfo{pages}{936--940} (\bibinfo{year}{2013}).

\bibitem{buttinoni2013dynamical}
\bibinfo{author}{Buttinoni, I.} \emph{et~al.}
\newblock \bibinfo{title}{Dynamical clustering and phase separation in
  suspensions of self-propelled colloidal particles}.
\newblock \emph{\bibinfo{journal}{Physical review letters}}
  \textbf{\bibinfo{volume}{110}}, \bibinfo{pages}{238301}
  (\bibinfo{year}{2013}).

\bibitem{van_der_linden_interrupted_2019}
\bibinfo{author}{van~der Linden, M.~N.}, \bibinfo{author}{Alexander, L.~C.},
  \bibinfo{author}{Aarts, D.~G.} \& \bibinfo{author}{Dauchot, O.}
\newblock \bibinfo{title}{Interrupted motility induced phase separation in
  aligning active colloids}.
\newblock \emph{\bibinfo{journal}{Physical Review Letters}}
  \textbf{\bibinfo{volume}{123}}, \bibinfo{pages}{098001}
  (\bibinfo{year}{2019}).

\bibitem{tan2022odd}
\bibinfo{author}{Tan, T.~H.} \emph{et~al.}
\newblock \bibinfo{title}{Odd dynamics of living chiral crystals}.
\newblock \emph{\bibinfo{journal}{Nature}} \textbf{\bibinfo{volume}{607}},
  \bibinfo{pages}{287--293} (\bibinfo{year}{2022}).

\bibitem{marchetti_hydrodynamics_2013}
\bibinfo{author}{Marchetti, M.~C.} \emph{et~al.}
\newblock \bibinfo{title}{Hydrodynamics of soft active matter}.
\newblock \emph{\bibinfo{journal}{Reviews of Modern Physics}}
  \textbf{\bibinfo{volume}{85}}, \bibinfo{pages}{1143--1189}
  (\bibinfo{year}{2013}).

\bibitem{soto_self-assembly_2014}
\bibinfo{author}{Soto, R.} \& \bibinfo{author}{Golestanian, R.}
\newblock \bibinfo{title}{Self-assembly of catalytically active colloidal
  molecules: Tailoring activity through surface chemistry}.
\newblock \emph{\bibinfo{journal}{Physical Review Letters}}
  \textbf{\bibinfo{volume}{112}}, \bibinfo{pages}{068301}
  (\bibinfo{year}{2014}).

\bibitem{baek2018generic}
\bibinfo{author}{Baek, Y.}, \bibinfo{author}{Solon, A.~P.},
  \bibinfo{author}{Xu, X.}, \bibinfo{author}{Nikola, N.} \&
  \bibinfo{author}{Kafri, Y.}
\newblock \bibinfo{title}{Generic long-range interactions between passive
  bodies in an active fluid}.
\newblock \emph{\bibinfo{journal}{Physical review letters}}
  \textbf{\bibinfo{volume}{120}}, \bibinfo{pages}{058002}
  (\bibinfo{year}{2018}).

\bibitem{saha_pairing_2019}
\bibinfo{author}{Saha, S.}, \bibinfo{author}{Ramaswamy, S.} \&
  \bibinfo{author}{Golestanian, R.}
\newblock \bibinfo{title}{Pairing, waltzing and scattering of chemotactic
  active colloids}.
\newblock \emph{\bibinfo{journal}{New Journal of Physics}}
  \textbf{\bibinfo{volume}{21}}, \bibinfo{pages}{063006}
  (\bibinfo{year}{2019}).

\bibitem{agudo-canalejo_active_2019}
\bibinfo{author}{Agudo-Canalejo, J.} \& \bibinfo{author}{Golestanian, R.}
\newblock \bibinfo{title}{Active phase separation in mixtures of chemically
  interacting particles}.
\newblock \emph{\bibinfo{journal}{Physical Review Letters}}
  \textbf{\bibinfo{volume}{123}}, \bibinfo{pages}{018101}
  (\bibinfo{year}{2019}).

\bibitem{saha_scalar_2020}
\bibinfo{author}{Saha, S.}, \bibinfo{author}{Agudo-Canalejo, J.} \&
  \bibinfo{author}{Golestanian, R.}
\newblock \bibinfo{title}{Scalar active mixtures: The nonreciprocal
  cahn-hilliard model}.
\newblock \emph{\bibinfo{journal}{Physical Review X}}
  \textbf{\bibinfo{volume}{10}}, \bibinfo{pages}{041009}
  (\bibinfo{year}{2020}).

\bibitem{you_nonreciprocity_2020}
\bibinfo{author}{You, Z.}, \bibinfo{author}{Baskaran, A.} \&
  \bibinfo{author}{Marchetti, M.~C.}
\newblock \bibinfo{title}{Nonreciprocity as a generic route to traveling
  states}.
\newblock \emph{\bibinfo{journal}{Proceedings of the National Academy of
  Sciences}} \textbf{\bibinfo{volume}{117}}, \bibinfo{pages}{19767--19772}
  (\bibinfo{year}{2020}).

\bibitem{nasouri_exact_2020}
\bibinfo{author}{Nasouri, B.} \& \bibinfo{author}{Golestanian, R.}
\newblock \bibinfo{title}{Exact phoretic interaction of two chemically active
  particles}.
\newblock \emph{\bibinfo{journal}{Physical Review Letters}}
  \textbf{\bibinfo{volume}{124}}, \bibinfo{pages}{168003}
  (\bibinfo{year}{2020}).

\bibitem{granek2020bodies}
\bibinfo{author}{Granek, O.}, \bibinfo{author}{Baek, Y.},
  \bibinfo{author}{Kafri, Y.} \& \bibinfo{author}{Solon, A.~P.}
\newblock \bibinfo{title}{Bodies in an interacting active fluid: far-field
  influence of a single body and interaction between two bodies}.
\newblock \emph{\bibinfo{journal}{Journal of Statistical Mechanics: Theory and
  Experiment}} \textbf{\bibinfo{volume}{2020}}, \bibinfo{pages}{063211}
  (\bibinfo{year}{2020}).

\bibitem{ouazan-reboul_non-equilibrium_2021}
\bibinfo{author}{Ouazan-Reboul, V.}, \bibinfo{author}{Agudo-Canalejo, J.} \&
  \bibinfo{author}{Golestanian, R.}
\newblock \bibinfo{title}{Non-equilibrium phase separation in mixtures of
  catalytically active particles: size dispersity and screening effects}.
\newblock \emph{\bibinfo{journal}{The European Physical Journal E}}
  \textbf{\bibinfo{volume}{44}}, \bibinfo{pages}{113} (\bibinfo{year}{2021}).

\bibitem{fruchart_non-rec_2021}
\bibinfo{author}{Fruchart, M.}, \bibinfo{author}{Hanai, R.},
  \bibinfo{author}{Littlewood, P.~B.} \& \bibinfo{author}{Vitelli, V.}
\newblock \bibinfo{title}{Non-reciprocal phase transitions}.
\newblock \emph{\bibinfo{journal}{Nature}} \textbf{\bibinfo{volume}{592}},
  \bibinfo{pages}{363--369} (\bibinfo{year}{2021}).

\bibitem{frohoff2021suppression}
\bibinfo{author}{Frohoff-H{\"u}lsmann, T.}, \bibinfo{author}{Wrembel, J.} \&
  \bibinfo{author}{Thiele, U.}
\newblock \bibinfo{title}{Suppression of coarsening and emergence of
  oscillatory behavior in a cahn-hilliard model with nonvariational coupling}.
\newblock \emph{\bibinfo{journal}{Physical Review E}}
  \textbf{\bibinfo{volume}{103}}, \bibinfo{pages}{042602}
  (\bibinfo{year}{2021}).

\bibitem{frohoff2021localized}
\bibinfo{author}{Frohoff-H{\"u}lsmann, T.} \& \bibinfo{author}{Thiele, U.}
\newblock \bibinfo{title}{Localized states in coupled cahn--hilliard
  equations}.
\newblock \emph{\bibinfo{journal}{IMA Journal of Applied Mathematics}}
  \textbf{\bibinfo{volume}{86}}, \bibinfo{pages}{924--943}
  (\bibinfo{year}{2021}).

\bibitem{poncet2022soft}
\bibinfo{author}{Poncet, A.} \& \bibinfo{author}{Bartolo, D.}
\newblock \bibinfo{title}{When soft crystals defy newton’s third law:
  Nonreciprocal mechanics and dislocation motility}.
\newblock \emph{\bibinfo{journal}{Physical Review Letters}}
  \textbf{\bibinfo{volume}{128}}, \bibinfo{pages}{048002}
  (\bibinfo{year}{2022}).

\bibitem{gupta2022nonreciprocal}
\bibinfo{author}{Gupta, R.~K.}, \bibinfo{author}{Kant, R.},
  \bibinfo{author}{Soni, H.}, \bibinfo{author}{Sood, A.~K.} \&
  \bibinfo{author}{Ramaswamy, S.}
\newblock \bibinfo{title}{Active nonreciprocal attraction between motile
  particles in an elastic medium}.
\newblock \emph{\bibinfo{journal}{Phys. Rev. E}}
  \textbf{\bibinfo{volume}{105}}, \bibinfo{pages}{064602}
  (\bibinfo{year}{2022}).

\bibitem{cross1993pattern}
\bibinfo{author}{Cross, M.~C.} \& \bibinfo{author}{Hohenberg, P.~C.}
\newblock \bibinfo{title}{Pattern formation outside of equilibrium}.
\newblock \emph{\bibinfo{journal}{Reviews of modern physics}}
  \textbf{\bibinfo{volume}{65}}, \bibinfo{pages}{851} (\bibinfo{year}{1993}).

\bibitem{saarloos_amplitude_1994}
\bibinfo{author}{Saarloos, W.}, \bibinfo{author}{Hecke, M.},
  \bibinfo{author}{Hohenberg, P.} \& \bibinfo{author}{Natuurwetenschappen, F.
  d. W.~e.}
\newblock \bibinfo{title}{Amplitude equations for pattern forming systems}
  (\bibinfo{year}{1994}).

\bibitem{aranson2002world}
\bibinfo{author}{Aranson, I.~S.} \& \bibinfo{author}{Kramer, L.}
\newblock \bibinfo{title}{The world of the complex ginzburg-landau equation}.
\newblock \emph{\bibinfo{journal}{Reviews of modern physics}}
  \textbf{\bibinfo{volume}{74}}, \bibinfo{pages}{99} (\bibinfo{year}{2002}).

\bibitem{rapp2019systematic}
\bibinfo{author}{Rapp, L.}, \bibinfo{author}{Bergmann, F.} \&
  \bibinfo{author}{Zimmermann, W.}
\newblock \bibinfo{title}{Systematic extension of the cahn-hilliard model for
  motility-induced phase separation}.
\newblock \emph{\bibinfo{journal}{The European Physical Journal E}}
  \textbf{\bibinfo{volume}{42}}, \bibinfo{pages}{1--10} (\bibinfo{year}{2019}).

\bibitem{bergmann2018active}
\bibinfo{author}{Bergmann, F.}, \bibinfo{author}{Rapp, L.} \&
  \bibinfo{author}{Zimmermann, W.}
\newblock \bibinfo{title}{Active phase separation: A universal approach}.
\newblock \emph{\bibinfo{journal}{Physical Review E}}
  \textbf{\bibinfo{volume}{98}}, \bibinfo{pages}{020603}
  (\bibinfo{year}{2018}).

\bibitem{stenhammar2015activity}
\bibinfo{author}{Stenhammar, J.}, \bibinfo{author}{Wittkowski, R.},
  \bibinfo{author}{Marenduzzo, D.} \& \bibinfo{author}{Cates, M.~E.}
\newblock \bibinfo{title}{Activity-induced phase separation and self-assembly
  in mixtures of active and passive particles}.
\newblock \emph{\bibinfo{journal}{Physical review letters}}
  \textbf{\bibinfo{volume}{114}}, \bibinfo{pages}{018301}
  (\bibinfo{year}{2015}).

\bibitem{yeo2015collective}
\bibinfo{author}{Yeo, K.}, \bibinfo{author}{Lushi, E.} \&
  \bibinfo{author}{Vlahovska, P.~M.}
\newblock \bibinfo{title}{Collective dynamics in a binary mixture of
  hydrodynamically coupled microrotors}.
\newblock \emph{\bibinfo{journal}{Physical review letters}}
  \textbf{\bibinfo{volume}{114}}, \bibinfo{pages}{188301}
  (\bibinfo{year}{2015}).

\bibitem{wysocki2016propagating}
\bibinfo{author}{Wysocki, A.}, \bibinfo{author}{Winkler, R.~G.} \&
  \bibinfo{author}{Gompper, G.}
\newblock \bibinfo{title}{Propagating interfaces in mixtures of active and
  passive brownian particles}.
\newblock \emph{\bibinfo{journal}{New journal of physics}}
  \textbf{\bibinfo{volume}{18}}, \bibinfo{pages}{123030}
  (\bibinfo{year}{2016}).

\bibitem{wittkowski2017nonequilibrium}
\bibinfo{author}{Wittkowski, R.}, \bibinfo{author}{Stenhammar, J.} \&
  \bibinfo{author}{Cates, M.~E.}
\newblock \bibinfo{title}{Nonequilibrium dynamics of mixtures of active and
  passive colloidal particles}.
\newblock \emph{\bibinfo{journal}{New Journal of Physics}}
  \textbf{\bibinfo{volume}{19}}, \bibinfo{pages}{105003}
  (\bibinfo{year}{2017}).

\bibitem{sturmer2019chemotaxis}
\bibinfo{author}{St{\"u}rmer, J.}, \bibinfo{author}{Seyrich, M.} \&
  \bibinfo{author}{Stark, H.}
\newblock \bibinfo{title}{Chemotaxis in a binary mixture of active and passive
  particles}.
\newblock \emph{\bibinfo{journal}{The Journal of chemical physics}}
  \textbf{\bibinfo{volume}{150}}, \bibinfo{pages}{214901}
  (\bibinfo{year}{2019}).

\bibitem{rodriguez2020phase}
\bibinfo{author}{Rodriguez, D.~R.}, \bibinfo{author}{Alarcon, F.},
  \bibinfo{author}{Martinez, R.}, \bibinfo{author}{Ram{\'\i}rez, J.} \&
  \bibinfo{author}{Valeriani, C.}
\newblock \bibinfo{title}{Phase behaviour and dynamical features of a
  two-dimensional binary mixture of active/passive spherical particles}.
\newblock \emph{\bibinfo{journal}{Soft Matter}} \textbf{\bibinfo{volume}{16}},
  \bibinfo{pages}{1162--1169} (\bibinfo{year}{2020}).

\bibitem{kolb2020active}
\bibinfo{author}{Kolb, T.} \& \bibinfo{author}{Klotsa, D.}
\newblock \bibinfo{title}{Active binary mixtures of fast and slow hard
  spheres}.
\newblock \emph{\bibinfo{journal}{Soft Matter}} \textbf{\bibinfo{volume}{16}},
  \bibinfo{pages}{1967--1978} (\bibinfo{year}{2020}).

\bibitem{bardfalvy2020symmetric}
\bibinfo{author}{B{\'a}rdfalvy, D.}, \bibinfo{author}{Anjum, S.},
  \bibinfo{author}{Nardini, C.}, \bibinfo{author}{Morozov, A.} \&
  \bibinfo{author}{Stenhammar, J.}
\newblock \bibinfo{title}{Symmetric mixtures of pusher and puller microswimmers
  behave as noninteracting suspensions}.
\newblock \emph{\bibinfo{journal}{Physical Review Letters}}
  \textbf{\bibinfo{volume}{125}}, \bibinfo{pages}{018003}
  (\bibinfo{year}{2020}).

\bibitem{de2021active}
\bibinfo{author}{de~Castro, P.}, \bibinfo{author}{Diles, S.},
  \bibinfo{author}{Soto, R.} \& \bibinfo{author}{Sollich, P.}
\newblock \bibinfo{title}{Active mixtures in a narrow channel: Motility
  diversity changes cluster sizes}.
\newblock \emph{\bibinfo{journal}{Soft Matter}} \textbf{\bibinfo{volume}{17}},
  \bibinfo{pages}{2050--2061} (\bibinfo{year}{2021}).

\bibitem{de2021diversity}
\bibinfo{author}{de~Castro, P.}, \bibinfo{author}{Rocha, F.~M.},
  \bibinfo{author}{Diles, S.}, \bibinfo{author}{Soto, R.} \&
  \bibinfo{author}{Sollich, P.}
\newblock \bibinfo{title}{Diversity of self-propulsion speeds reduces
  motility-induced clustering in confined active matter}.
\newblock \emph{\bibinfo{journal}{Soft Matter}} \textbf{\bibinfo{volume}{17}},
  \bibinfo{pages}{9926--9936} (\bibinfo{year}{2021}).

\bibitem{paoluzzi2020information}
\bibinfo{author}{Paoluzzi, M.}, \bibinfo{author}{Leoni, M.} \&
  \bibinfo{author}{Marchetti, M.~C.}
\newblock \bibinfo{title}{Information and motility exchange in collectives of
  active particles}.
\newblock \emph{\bibinfo{journal}{Soft Matter}} \textbf{\bibinfo{volume}{16}},
  \bibinfo{pages}{6317--6327} (\bibinfo{year}{2020}).

\bibitem{li2021hierarchical}
\bibinfo{author}{Li, Y.~I.} \& \bibinfo{author}{Cates, M.~E.}
\newblock \bibinfo{title}{Hierarchical microphase separation in non-conserved
  active mixtures}.
\newblock \emph{\bibinfo{journal}{The European Physical Journal E}}
  \textbf{\bibinfo{volume}{44}}, \bibinfo{pages}{1--8} (\bibinfo{year}{2021}).

\bibitem{williams2021confinement}
\bibinfo{author}{Williams, S.}, \bibinfo{author}{Jeanneret, R.},
  \bibinfo{author}{Tuval, I.} \& \bibinfo{author}{Polin, M.}
\newblock \bibinfo{title}{Confinement-induced accumulation and spontaneous
  de-mixing of microscopic active-passive mixtures}.
\newblock \emph{\bibinfo{journal}{arXiv preprint arXiv:2111.09763}}
  (\bibinfo{year}{2021}).

\bibitem{miller2001QS}
\bibinfo{author}{Miller, M.~B.} \& \bibinfo{author}{Bassler, B.~L.}
\newblock \bibinfo{title}{Quorum sensing in bacteria}.
\newblock \emph{\bibinfo{journal}{Annual Review of Microbiology}}
  \textbf{\bibinfo{volume}{55}}, \bibinfo{pages}{165--199}
  (\bibinfo{year}{2001}).

\bibitem{nealson1970luminescence}
\bibinfo{author}{Nealson, K.~H.}, \bibinfo{author}{Platt, T.} \&
  \bibinfo{author}{Hastings, J.~W.}
\newblock \bibinfo{title}{Cellular control of the synthesis and activity of the
  bacterial luminescent system}.
\newblock \emph{\bibinfo{journal}{Journal of Bacteriology}}
  \textbf{\bibinfo{volume}{104}}, \bibinfo{pages}{313--322}
  (\bibinfo{year}{1970}).

\bibitem{engebrecht1984lux}
\bibinfo{author}{Engebrecht, J.} \& \bibinfo{author}{Silverman, M.}
\newblock \bibinfo{title}{Identification of genes and gene products necessary
  for bacterial bioluminescence.}
\newblock \emph{\bibinfo{journal}{Proceedings of the National Academy of
  Sciences}} \textbf{\bibinfo{volume}{81}}, \bibinfo{pages}{4154--4158}
  (\bibinfo{year}{1984}).

\bibitem{fuqua1994quorum}
\bibinfo{author}{Fuqua, W.~C.}, \bibinfo{author}{Winans, S.~C.} \&
  \bibinfo{author}{Greenberg, E.~P.}
\newblock \bibinfo{title}{Quorum sensing in bacteria: the luxr-luxi family of
  cell density-responsive transcriptional regulators}.
\newblock \emph{\bibinfo{journal}{Journal of bacteriology}}
  \textbf{\bibinfo{volume}{176}}, \bibinfo{pages}{269--275}
  (\bibinfo{year}{1994}).

\bibitem{verma2013quorum}
\bibinfo{author}{Verma, S.~C.} \& \bibinfo{author}{Miyashiro, T.}
\newblock \bibinfo{title}{Quorum sensing in the squid-vibrio symbiosis}.
\newblock \emph{\bibinfo{journal}{International journal of molecular sciences}}
  \textbf{\bibinfo{volume}{14}}, \bibinfo{pages}{16386--16401}
  (\bibinfo{year}{2013}).

\bibitem{tsou2010virulence}
\bibinfo{author}{Tsou, A.~M.} \& \bibinfo{author}{Zhu, J.}
\newblock \bibinfo{title}{Quorum sensing negatively regulates hemolysin
  transcriptionally and posttranslationally in <i>vibrio cholerae</i>}.
\newblock \emph{\bibinfo{journal}{Infection and Immunity}}
  \textbf{\bibinfo{volume}{78}}, \bibinfo{pages}{461--467}
  (\bibinfo{year}{2010}).

\bibitem{hammer2003biofilm}
\bibinfo{author}{Hammer, B.~K.} \& \bibinfo{author}{Bassler, B.~L.}
\newblock \bibinfo{title}{Quorum sensing controls biofilm formation in vibrio
  cholerae}.
\newblock \emph{\bibinfo{journal}{Molecular Microbiology}}
  \textbf{\bibinfo{volume}{50}}, \bibinfo{pages}{101--104}
  (\bibinfo{year}{2003}).

\bibitem{daniels2004swarming}
\bibinfo{author}{Daniels, R.}, \bibinfo{author}{Vanderleyden, J.} \&
  \bibinfo{author}{Michiels, J.}
\newblock \bibinfo{title}{{Quorum sensing and swarming migration in bacteria}}.
\newblock \emph{\bibinfo{journal}{FEMS Microbiology Reviews}}
  \textbf{\bibinfo{volume}{28}}, \bibinfo{pages}{261--289}
  (\bibinfo{year}{2004}).

\bibitem{bauerle_self-organization_2018}
\bibinfo{author}{Bäuerle, T.}, \bibinfo{author}{Fischer, A.},
  \bibinfo{author}{Speck, T.} \& \bibinfo{author}{Bechinger, C.}
\newblock \bibinfo{title}{Self-organization of active particles by quorum
  sensing rules}.
\newblock \emph{\bibinfo{journal}{Nature Communications}}
  \textbf{\bibinfo{volume}{9}}, \bibinfo{pages}{3232} (\bibinfo{year}{2018}).

\bibitem{lavergne_group_2019}
\bibinfo{author}{Lavergne, F.~A.}, \bibinfo{author}{Wendehenne, H.},
  \bibinfo{author}{Bäuerle, T.} \& \bibinfo{author}{Bechinger, C.}
\newblock \bibinfo{title}{Group formation and cohesion of active particles with
  visual perception–dependent motility}.
\newblock \emph{\bibinfo{journal}{Science}}  (\bibinfo{year}{2019}).

\bibitem{massana-cid_rectification_2022}
\bibinfo{author}{Massana-Cid, H.}, \bibinfo{author}{Maggi, C.},
  \bibinfo{author}{Frangipane, G.} \& \bibinfo{author}{Di~Leonardo, R.}
\newblock \bibinfo{title}{Rectification and confinement of photokinetic
  bacteria in an optical feedback loop}.
\newblock \emph{\bibinfo{journal}{Nature Communications}}
  \textbf{\bibinfo{volume}{13}}, \bibinfo{pages}{2740} (\bibinfo{year}{2022}).

\bibitem{Note1}
\bibinfo{note}{Note that making $\tau _{i,\mu }$ a function of ${\protect \bf
  r}_i$ and a functional of $\{\rho _\nu \}$ does not lead to any interesting
  phenomenology so, for simplicity, we do not consider this case in the main
  text.}

\bibitem{hohenberg_theory_1977}
\bibinfo{author}{Hohenberg, P.~C.} \& \bibinfo{author}{Halperin, B.~I.}
\newblock \bibinfo{title}{Theory of dynamic critical phenomena}.
\newblock \emph{\bibinfo{journal}{Reviews of Modern Physics}}
  \textbf{\bibinfo{volume}{49}}, \bibinfo{pages}{435--479}
  (\bibinfo{year}{1977}).

\bibitem{curatolo_cooperative_2020}
\bibinfo{author}{Curatolo, A.~I.} \emph{et~al.}
\newblock \bibinfo{title}{Cooperative pattern formation in multi-component
  bacterial systems through reciprocal motility regulation}.
\newblock \emph{\bibinfo{journal}{Nature Physics}}
  \textbf{\bibinfo{volume}{16}}, \bibinfo{pages}{1152--1157}
  (\bibinfo{year}{2020}).

\bibitem{obyrne_lamellar_2020}
\bibinfo{author}{O’Byrne, J.} \& \bibinfo{author}{Tailleur, J.}
\newblock \bibinfo{title}{Lamellar to micellar phases and beyond: When tactic
  active systems admit free energy functionals}.
\newblock \emph{\bibinfo{journal}{Physical Review Letters}}
  \textbf{\bibinfo{volume}{125}}, \bibinfo{pages}{208003}
  (\bibinfo{year}{2020}).

\bibitem{cates_motility-induced_2015}
\bibinfo{author}{Cates, M.~E.} \& \bibinfo{author}{Tailleur, J.}
\newblock \bibinfo{title}{Motility-induced phase separation}.
\newblock \emph{\bibinfo{journal}{Annual Review of Condensed Matter Physics}}
  \textbf{\bibinfo{volume}{6}}, \bibinfo{pages}{219--244}
  (\bibinfo{year}{2015}).

\bibitem{fodor2018statistical}
\bibinfo{author}{Fodor, {\'E}.} \& \bibinfo{author}{Marchetti, M.~C.}
\newblock \bibinfo{title}{The statistical physics of active matter: From
  self-catalytic colloids to living cells}.
\newblock \emph{\bibinfo{journal}{Physica A: Statistical Mechanics and its
  Applications}} \textbf{\bibinfo{volume}{504}}, \bibinfo{pages}{106--120}
  (\bibinfo{year}{2018}).

\bibitem{sollich_predicting_2001}
\bibinfo{author}{Sollich, P.}
\newblock \bibinfo{title}{Predicting phase equilibria in polydisperse systems}.
\newblock \emph{\bibinfo{journal}{Journal of Physics: Condensed Matter}}
  \textbf{\bibinfo{volume}{14}}, \bibinfo{pages}{R79--R117}
  (\bibinfo{year}{2001}).

\bibitem{cross_pattern_1993}
\bibinfo{author}{Cross, M.~C.} \& \bibinfo{author}{Hohenberg, P.~C.}
\newblock \bibinfo{title}{Pattern formation outside of equilibrium}.
\newblock \emph{\bibinfo{journal}{Reviews of Modern Physics}}
  \textbf{\bibinfo{volume}{65}}, \bibinfo{pages}{851--1112}
  (\bibinfo{year}{1993}).

\bibitem{schnitzer_theory_1993}
\bibinfo{author}{Schnitzer, M.~J.}
\newblock \bibinfo{title}{Theory of continuum random walks and application to
  chemotaxis}.
\newblock \emph{\bibinfo{journal}{Phys. Rev. E}} \textbf{\bibinfo{volume}{48}},
  \bibinfo{pages}{2553--2568} (\bibinfo{year}{1993}).

\bibitem{tailleur_statistical_2008}
\bibinfo{author}{Tailleur, J.} \& \bibinfo{author}{Cates, M.~E.}
\newblock \bibinfo{title}{Statistical mechanics of interacting run-and-tumble
  bacteria}.
\newblock \emph{\bibinfo{journal}{Physical Review Letters}}
  \textbf{\bibinfo{volume}{100}}, \bibinfo{pages}{218103}
  (\bibinfo{year}{2008}).

\bibitem{cates_when_2013}
\bibinfo{author}{Cates, M.~E.} \& \bibinfo{author}{Tailleur, J.}
\newblock \bibinfo{title}{When are active brownian particles and run-and-tumble
  particles equivalent? consequences for motility-induced phase separation}.
\newblock \emph{\bibinfo{journal}{{EPL} - Europhysics Letters}}
  \textbf{\bibinfo{volume}{101}}, \bibinfo{pages}{20010}
  (\bibinfo{year}{2013}).

\bibitem{martin_PRE_2021}
\bibinfo{author}{Martin, D.} \emph{et~al.}
\newblock \bibinfo{title}{Statistical mechanics of active ornstein-uhlenbeck
  particles}.
\newblock \emph{\bibinfo{journal}{Phys. Rev. E}}
  \textbf{\bibinfo{volume}{103}}, \bibinfo{pages}{032607}
  (\bibinfo{year}{2021}).

\bibitem{frangipane_dynamic_2018}
\bibinfo{author}{Frangipane, G.} \emph{et~al.}
\newblock \bibinfo{title}{Dynamic density shaping of photokinetic e. coli}.
\newblock \emph{\bibinfo{journal}{{eLife}}} \textbf{\bibinfo{volume}{7}},
  \bibinfo{pages}{e36608} (\bibinfo{year}{2018}).

\bibitem{arlt_painting_2018}
\bibinfo{author}{Arlt, J.}, \bibinfo{author}{Martinez, V.~A.},
  \bibinfo{author}{Dawson, A.}, \bibinfo{author}{Pilizota, T.} \&
  \bibinfo{author}{Poon, W. C.~K.}
\newblock \bibinfo{title}{Painting with light-powered bacteria}.
\newblock \emph{\bibinfo{journal}{Nat Commun}} \textbf{\bibinfo{volume}{9}},
  \bibinfo{pages}{768} (\bibinfo{year}{2018}).

\bibitem{berg1975chemotaxis}
\bibinfo{author}{Berg, H.~C.}
\newblock \bibinfo{title}{Chemotaxis in bacteria}.
\newblock \emph{\bibinfo{journal}{Annual review of biophysics and
  bioengineering}} \textbf{\bibinfo{volume}{4}}, \bibinfo{pages}{119--136}
  (\bibinfo{year}{1975}).

\bibitem{budrene1995dynamics}
\bibinfo{author}{Budrene, E.~O.} \& \bibinfo{author}{Berg, H.~C.}
\newblock \bibinfo{title}{Dynamics of formation of symmetrical patterns by
  chemotactic bacteria}.
\newblock \emph{\bibinfo{journal}{Nature}} \textbf{\bibinfo{volume}{376}},
  \bibinfo{pages}{49--53} (\bibinfo{year}{1995}).

\bibitem{woodward1995spatio}
\bibinfo{author}{Woodward, D.~E.} \emph{et~al.}
\newblock \bibinfo{title}{Spatio-temporal patterns generated by salmonella
  typhimurium}.
\newblock \emph{\bibinfo{journal}{Biophysical journal}}
  \textbf{\bibinfo{volume}{68}}, \bibinfo{pages}{2181--2189}
  (\bibinfo{year}{1995}).

\bibitem{brenner1998physical}
\bibinfo{author}{Brenner, M.~P.}, \bibinfo{author}{Levitov, L.~S.} \&
  \bibinfo{author}{Budrene, E.~O.}
\newblock \bibinfo{title}{Physical mechanisms for chemotactic pattern formation
  by bacteria}.
\newblock \emph{\bibinfo{journal}{Biophysical journal}}
  \textbf{\bibinfo{volume}{74}}, \bibinfo{pages}{1677--1693}
  (\bibinfo{year}{1998}).

\bibitem{saragosti2011directional}
\bibinfo{author}{Saragosti, J.} \emph{et~al.}
\newblock \bibinfo{title}{Directional persistence of chemotactic bacteria in a
  traveling concentration wave}.
\newblock \emph{\bibinfo{journal}{Proceedings of the National Academy of
  Sciences}} \textbf{\bibinfo{volume}{108}}, \bibinfo{pages}{16235--16240}
  (\bibinfo{year}{2011}).

\bibitem{chatterjee2011chemotaxis}
\bibinfo{author}{Chatterjee, S.}, \bibinfo{author}{da~Silveira, R.~A.} \&
  \bibinfo{author}{Kafri, Y.}
\newblock \bibinfo{title}{Chemotaxis when bacteria remember: drift versus
  diffusion}.
\newblock \emph{\bibinfo{journal}{PLoS computational biology}}
  \textbf{\bibinfo{volume}{7}}, \bibinfo{pages}{e1002283}
  (\bibinfo{year}{2011}).

\bibitem{sourjik2012responding}
\bibinfo{author}{Sourjik, V.} \& \bibinfo{author}{Wingreen, N.~S.}
\newblock \bibinfo{title}{Responding to chemical gradients: bacterial
  chemotaxis}.
\newblock \emph{\bibinfo{journal}{Current opinion in cell biology}}
  \textbf{\bibinfo{volume}{24}}, \bibinfo{pages}{262--268}
  (\bibinfo{year}{2012}).

\bibitem{cremer2019chemotaxis}
\bibinfo{author}{Cremer, J.} \emph{et~al.}
\newblock \bibinfo{title}{Chemotaxis as a navigation strategy to boost range
  expansion}.
\newblock \emph{\bibinfo{journal}{Nature}} \textbf{\bibinfo{volume}{575}},
  \bibinfo{pages}{658--663} (\bibinfo{year}{2019}).

\bibitem{newman2004many}
\bibinfo{author}{Newman, T.} \& \bibinfo{author}{Grima, R.}
\newblock \bibinfo{title}{Many-body theory of chemotactic cell-cell
  interactions}.
\newblock \emph{\bibinfo{journal}{Physical Review E}}
  \textbf{\bibinfo{volume}{70}}, \bibinfo{pages}{051916}
  (\bibinfo{year}{2004}).

\bibitem{chavanis2007exact}
\bibinfo{author}{Chavanis, P.-H.}
\newblock \bibinfo{title}{Exact diffusion coefficient of self-gravitating
  brownian particles in two dimensions}.
\newblock \emph{\bibinfo{journal}{The European Physical Journal B}}
  \textbf{\bibinfo{volume}{57}}, \bibinfo{pages}{391--409}
  (\bibinfo{year}{2007}).

\bibitem{cates2010arrested}
\bibinfo{author}{Cates, M.~E.}, \bibinfo{author}{Marenduzzo, D.},
  \bibinfo{author}{Pagonabarraga, I.} \& \bibinfo{author}{Tailleur, J.}
\newblock \bibinfo{title}{Arrested phase separation in reproducing bacteria
  creates a generic route to pattern formation}.
\newblock \emph{\bibinfo{journal}{Proceedings of the National Academy of
  Sciences}} \textbf{\bibinfo{volume}{107}}, \bibinfo{pages}{11715--11720}
  (\bibinfo{year}{2010}).

\bibitem{liu_sequential_2011}
\bibinfo{author}{Liu, C.} \emph{et~al.}
\newblock \bibinfo{title}{Sequential establishment of stripe patterns in an
  expanding cell population}.
\newblock \emph{\bibinfo{journal}{Science}}  (\bibinfo{year}{2011}).

\bibitem{hallatschek_genetic_2007}
\bibinfo{author}{Hallatschek, O.}, \bibinfo{author}{Hersen, P.},
  \bibinfo{author}{Ramanathan, S.} \& \bibinfo{author}{Nelson, D.~R.}
\newblock \bibinfo{title}{Genetic drift at expanding frontiers promotes gene
  segregation}.
\newblock \emph{\bibinfo{journal}{Proceedings of the National Academy of
  Sciences}} \textbf{\bibinfo{volume}{104}}, \bibinfo{pages}{19926--19930}
  (\bibinfo{year}{2007}).

\bibitem{goodrich2021designing}
\bibinfo{author}{Goodrich, C.~P.}, \bibinfo{author}{King, E.~M.},
  \bibinfo{author}{Schoenholz, S.~S.}, \bibinfo{author}{Cubuk, E.~D.} \&
  \bibinfo{author}{Brenner, M.~P.}
\newblock \bibinfo{title}{Designing self-assembling kinetics with
  differentiable statistical physics models}.
\newblock \emph{\bibinfo{journal}{Proceedings of the National Academy of
  Sciences}} \textbf{\bibinfo{volume}{118}}, \bibinfo{pages}{e2024083118}
  (\bibinfo{year}{2021}).

\end{thebibliography}
\end{document}